\documentclass[showpacs,preprintnumbers,twocolumn,reprint,superscriptaddress,nobibnotes,nofootinbib]{revtex4-1}

\usepackage{amsmath,amssymb,hyperref,times,graphicx,bm,subfigure}
\def\<{\langle}
\def\>{\rangle}
\DeclareMathOperator{\Tr}{Tr}
\newcommand{\PA}{Physica}
\newcommand{\PAA}{\PA~A}

\newcommand{\PR}{Phys.~Rev.}
\newcommand{\PRL}{Phys.~Rev.~Lett.}

\newcommand{\PRE}{Phys.~Rev. E}

\newcommand{\JPAOLD}{J.~Phys.~A: Math.~Gen.}

\newcommand{\physrep}{Phys.~Rep.}

\newcommand{\ZPC}{Z.~Phys.~C}

\newcommand{\NPB}{Nucl.~Phys.~B}
\newcommand{\JSF}{J.~Stat.~Phys.}
\newcommand{\PLB}{Phys.~Lett.~B}

\begin{document}

\title{Finite-size effects in canonical and grand-canonical quantum Monte Carlo simulations for fermions}

\author{\firstname{Zhenjiu} \surname{Wang}}
\email{Zhenjiu.Wang@physik.uni-wuerzburg.de}
\author{\firstname{Fakher F.} \surname{Assaad}}
\email{assaad@physik.uni-wuerzburg.de}
\author{\firstname{Francesco} \surname{Parisen Toldin}}
\email{francesco.parisentoldin@physik.uni-wuerzburg.de}
\affiliation{\mbox{Institut f\"ur Theoretische Physik und Astrophysik, Universit\"at W\"urzburg, Am Hubland, D-97074 W\"urzburg, Germany}}
\begin{abstract}
We introduce a quantum Monte Carlo method at finite temperature for interacting fermionic models in the canonical ensemble, where the conservation of the particle number is enforced. Although general thermodynamic arguments ensure the equivalence of the canonical and the grand-canonical ensembles in the thermodynamic limit, their approach to the infinite-volume limit is distinctively different. Observables computed in the canonical ensemble generically display a finite-size correction proportional to the inverse volume, whereas in the grand-canonical ensemble the approach is exponential in the ratio of the linear size over the correlation length. We verify these predictions by quantum Monte Carlo simulations of the Hubbard model in one and two dimensions in the grand-canonical and the canonical ensemble. We prove an exact formula for the finite-size part of the free energy density, energy density and other observables in the canonical ensemble and relate this correction to a susceptibility computed in the corresponding grand-canonical ensemble. This result is confirmed by an exact computation of the one-dimensional classical Ising model in the canonical ensemble, which for classical models corresponds to the so-called fixed-magnetization ensemble. Our method is useful for simulating finite systems which are not coupled to a particle bath, such as in nuclear or cold atom physics.
\end{abstract}

\maketitle

\section{Introduction}
\label{sec:intro}

One of the central tenet of statistical mechanics is the notion of statistical ensembles. In thermal equilibrium, a system can be described by different statistical ensembles: the microcanonical, the canonical, and the grand-canonical ensemble. In the thermodynamic limit, and in the presence of short-ranged interactions, bulk properties do not generically depend on the choice of the ensemble. Such a property is known as ensemble equivalence.\footnote{Nevertheless, systems with long-ranged interactions exhibit violation of the ensemble equivalence \cite{BMR-01,LR-02}.} In particular, a textbook argument for the equivalence between the canonical and the grand-canonical ensembles consists in the following observation. In the grand-canonical ensemble the particle number as well as the energy are \textit{sharp}  in the  thermodynamic limit, i.e., their relative fluctuation vanishes in the thermodynamic limit. This stems from the fact that  the specific heat, 
\begin{equation}
	C_V   = \frac{d \langle \hat{H}  \rangle}{d T}    = k_B\beta^2  \left(  \langle \hat{H}^2 \rangle - \langle \hat{H} \rangle^2  \right),
\label{specific_heat}
\end{equation}
and  the charge susceptibility
\begin{equation}
	\Xi_c  =  \frac{d \langle \hat{N}  \rangle}{d \mu }    =  \beta \left(  \langle \hat{N}^2 \rangle - \langle \hat{N} \rangle^2  \right),
\label{charge_chi_extensive}
\end{equation}
are extensive quantities  that measure  energy and particle-number fluctuations.
 In Eqs.~(\ref{specific_heat}) and (\ref{charge_chi_extensive}) $\hat{H}$ is the Hamiltonian of the system, $\hat{N}$ the particle-number operator, $\beta=1/k_BT$ the inverse temperature in units of the Boltzmann constant $k_B$, and $\mu$ the chemical potential.
 Thus, 
\begin{equation}
	 \lim_{N \rightarrow \infty} \frac{ \sqrt{  \left(  \langle \hat{H}^2 \rangle - \langle \hat{H} \rangle^2  \right) }} {  \langle \hat{H} \rangle }  = 
	 \lim_{N \rightarrow \infty} \frac{ \sqrt{  \left(  \langle \hat{N}^2 \rangle - \langle \hat{N} \rangle^2  \right) }} {  \langle \hat{N} \rangle } = 0
\end{equation}
and the selection of the ensemble is merely a matter of convenience.
  Nonetheless, in many cases the choice of ensemble is dictated by the physical properties of the system under study. In fact, while the canonical ensemble requires the presence of a heat bath which fixes the temperature, the grand-canonical ensemble additionally needs a particle reservoir which allows to fix the chemical potential. Systems which lack such a particle bath, like those found in nuclear physics or in cold atoms, require a description in terms of the canonical ensemble. Moreover, in the case of mesoscopic systems with a finite particle number, a reliable comparison with experimental data needs a theoretical computation based on the canonical ensemble. In this context, we mention that, unlike the finite-temperature auxiliary field quantum Monte Carlo (QMC) method considered here, the so-called projective auxiliary field QMC, which targets the ground state of fermionic models, is a method which is intrinsically formulated within the canonical ensemble \cite{Assaad08_rev}.

The aim of this  paper is twofold.  On one hand we will introduce a QMC method for fermionic models in the canonical ensemble, consisting in a simple formulation of the auxiliary field QMC method which enforces the conservation of the particle number. Our approach differs from that  adopted in Ref.~\cite{KDK-97}  and supplements the Hamiltonian that we simulate in the grand-canonical ensemble by the long-ranged interaction term
\begin{equation}
     \lambda \left( \hat{N} - N \right)^2,
\end{equation} 
such that in  the infinite-$\lambda$ limit charge fluctuations are suppressed and the canonical ensemble is recovered.   This type of  interaction is easily  incorporated in the  auxiliary field QMC, especially in the formulation provided in Ref.~\cite{ALF}.   The advantage of such an approach is that $ \lambda$ can be dynamically chosen.  For instance, at low temperatures  the charge susceptibility  can  vanish due to finite size  or correlation-induced charge gaps. In this case $\lambda$ can be set to a very small number, or even to zero since both  canonical and grand-canonical ensembles  yield identical results.
 At high temperatures, where the grand-canonical ensemble exhibits significantly large  charge fluctuations, bigger  values of $\lambda $ are required to impose the constraint. 

The second motivation of the paper is to look into finite-size corrections both in the canonical and grand-canonical ensembles, which we study in quantum and classical lattice models.

Concerning classical models on a lattice, it should be noted that in the literature the canonical ensemble is often defined by the usual partition function sum, where one considers all the configurations without any constraint. In the case of the standard Ising model, this corresponds to the usual partition function:
\begin{equation}
Z_{\rm gc}(h) = \sum_{\{S_k=\pm 1\}}\exp\left\{\beta J\sum_{<i j>}S_iS_j+h\sum_i S_i\right\}.
\label{ZIsing_gc}
\end{equation}
However, through the mapping to the lattice gas, the magnetization of the model corresponds to the particle number, which in the ensemble of Eq.~(\ref{ZIsing_gc}) is allowed to fluctuate. In order to provide a more meaningful comparison to quantum models, we refer to the lattice gas language and define the grand-canonical ensemble as the one where the magnetization is not fixed; in Eq.~(\ref{ZIsing_gc}) we have anticipated this definition, such that the subscript gc refers the grand-canonical ensemble. Conversely, we define the canonical ensemble as the ensemble where the magnetization is fixed, so that the corresponding partition function of the Ising model is
\begin{equation}
\begin{split}
Z_{\rm can}(h,m) = \sum_{\{S_k=\pm 1\}}&\exp\left\{\beta J\sum_{<i j>}S_iS_j+h\sum_i S_i\right\}\\
&\ \ \cdot\delta\left(m, \frac{1}{V}\sum_i S_i\right),
\end{split}
\label{ZIsing_can}
\end{equation}
where the constraint is enforced by employing the Kronecker delta function $\delta(m,n)$. In the literature, the ensemble of Eq.~(\ref{ZIsing_can}) is often referred to as the {\it fixed-magnetization} ensemble. In three dimensions, the Ising model at fixed magnetization has been investigated by means of Monte Carlo (MC) simulations in Ref.~\cite{BHT-00}, using the geometric cluster algorithm \cite{HB-98,HB-98B}.

In this work, we study the approach to the thermodynamic limit in the presence of a finite mass gap or, in the language of statistical physics, with a finite exponential correlation length. Generically for short-ranged Hamiltonians, on a finite volume with periodic boundary conditions and in the grand-canonical ensemble, the various observables are expected to show a finite-size correction which is proportional to $\exp(-L/\xi)$, where $L$ is the linear size of the system and $\xi$ is the exponential correlation length (or inverse mass gap). This expectation has been confirmed by explicit field theory calculations, both in the continuum \cite{Luescher-86,Neuberger-89} and on a lattice \cite{Muenster-85}; early numerical studies confirmed these prediction \cite{MW-87}. An exponential approach to the thermodynamic limit is also verified, e.g., in the well-known solution of the one-dimensional Ising model, as well as in generic one-dimensional $O(N)-$invariant spin models \cite{CMPS-97}. Nevertheless, it should be noted that, in the grand-canonical ensemble, some specific observables can exhibit a leading finite-size correction proportional to a power law of the system size. This is the case of the most common definition of the second-moment correlation length on a lattice, where finite-size corrections $\propto 1/L^2$ are due to the discretization of momenta on a finite lattice; see, e.g., the corresponding discussion in Ref.~\cite{CP-98} and Appendix A of Ref.~\cite{PTHAH-14}. We also remark that, in the presence of nontranslationally invariant boundary conditions, finite-size corrections polynomial in the inverse lattice size $1/L$ arise naturally, being related to subleading terms in the free energy; for instance, open boundary conditions result in the presence of a surface free energy which is depressed by a factor $1/L$ with respect to the bulk one and gives rise to finite-size corrections $\propto 1/L$ for bulk observables.

Conversely, in the canonical ensemble the prediction of exponentially decaying finite-size corrections fails, since the constraint introduces a long-ranged interaction, such that fluctuations in spatially separated regions (as measured by the correlation length) are not independent. Such a long-ranged (weak) correlation modifies also the high-temperature expansion of a model \cite{ISR-15} and results in a slower approach to the thermodynamic limit of various observables, so that the leading finite-size correction is proportional to the inverse volume $V$. Several important properties of the free energy in the canonical ensemble have been, in fact, discussed in the literature, although under a different perspective and notation. In quantum field theory, the so-called constrained effective potential $U_{\rm eff}$, introduced in the context of scalar field theories in Ref.~\cite{FK-75}, is defined as
\begin{equation}
e^{-VU_{\rm eff}(m,V)} = \int [{\cal D}\varphi] e^{-S[\varphi]}\delta\left(m - \frac{1}{V}\int d^dx\varphi(x)\right),
\label{Ueff_def}
\end{equation}
where $S[\varphi]$ is the action of the theory and the right-hand side of Eq.~(\ref{Ueff_def}) is a constrained path-integral over the field configurations where the volume-average value of $\varphi$ is fixed to $m$. In the language of statistical physics, the right-hand side of Eq.~(\ref{Ueff_def}) is precisely a constrained partition function sum at fixed magnetization, i.e., the partition function in the canonical ensemble. Hence, $U_{\rm eff}(m,V)$ is the free energy per volume and $k_BT$ in the canonical ensemble. A detailed analysis of the constrained effective potential has shown that it admits an infinite-volume limit $U_{\rm eff}(m,V\rightarrow\infty)$ which coincides with the usual effective potential $\Gamma(m)$ of the theory \cite{ORWY-86}. Moreover, as argued in Ref.~\cite{Neuberger-89}, $U_{\rm eff}(m,V)$ exhibits finite-size corrections which are polynomial in $1/V$. This is because, as a consequence of the definition in Eq.~(\ref{Ueff_def}), the grand-canonical average of any function of the magnetization $m$ is equivalent to an average over an effective probability measure $\propto \exp\{-VU_{\rm eff}(m,V)\}$, which for $V\rightarrow\infty$ can be evaluated by a saddle-point expansion, resulting in a series in $1/V$. On the other hand, the grand-canonical average converges exponentially to the limit $V\rightarrow\infty$. This is possible only if $U_{\rm eff}(m,V)$ displays finite-size corrections polynomial in $1/V$, which exactly cancel the expansion in $1/V$ originating from the saddle-point evaluation \cite{Neuberger-89}. A renormalized loop expansion for a $\phi^4$ theory on the lattice has confirmed the existence of finite-size corrections $\propto 1/V$ \cite{Palma-92}.

In this context, a recent study verified the existence of finite-size corrections $\propto 1/V$ in the canonical ensemble, and, conversely, of exponentially decaying finite-size corrections in the grand-canonical ensemble \cite{ISR-15}.
In this paper we provide an exact formula for the leading finite-size corrections in the canonical ensemble of the free energy density, energy density and other observables. While our analysis is restricted to the case of a finite correlation length, we mention that the introduction of a constraint to a nonordering parameter results in the so-called Fisher renormalization, leading to a modification of the singularities associated with a critical point, such that the critical exponents differ from those observed in the corresponding unconstrained system \cite{EG-67,Fisher-68}. The choice of ensemble is also relevant to the so-called critical Casimir force \cite{FG-78}, whose behavior in the canonical ensemble has been recently investigated within mean-field theory and MC simulations~\cite{GVGD-16}.

This paper is organized as follows. In Sec.~\ref{sec:method} we illustrate the QMC method that we use to generate numerical data for fermionic models in the canonical ensemble. In Sec.~\ref{sec:fscanonical} we provide an exact determination of the leading finite-size corrections in the canonical ensemble. In Sec.~\ref{sec:qmc} we study the finite-size corrections of the Hubbard model in one and two dimensions. In Sec.~\ref{sec:conclusions} we summarize our results. In the Appendix we provide an exact solution of the one-dimensional classical Ising model in the canonical ensemble to the leading order in $1/V$, which confirms the general result of Sec.~\ref{sec:fscanonical}.
\section{Canonical auxiliary field methods}
\label{sec:method}
\subsection{General formulation}
\label{sec:method:general}

In this section  we review  various methods to achieve canonical  auxiliary field  QMC simulations at  finite temperature.   We will consider a Hamiltonian  of the form
\begin{equation}
\begin{split}
  &\hat{H}  = \hat{T} + \hat{V},\\
  &\hat{T}  \equiv  \sum_{x,y} \hat{c}^{\dagger}_{x}  T_{x,y} \hat{c}^{\phantom\dagger}_{y},\\
  &\hat{V} \equiv \sum_{k} U_{k} \left(\hat{V}^{(k)} +  \alpha_k \right)^2,\qquad  \hat{V}^{(k)}\equiv  \sum_{x,y} \hat{c}^{\dagger}_{x}  V^{(k)}_{x,y} \hat{c}^{\phantom\dagger}_{y},
\end{split}
  \end{equation}
that can be readily implemented in the ALF package \cite{ALF}.   Here $x$ is a super-index encoding orbital and spin degrees of freedom, $\hat{c}^{\dagger}_{x}$  are fermion creation operators, $V^{(k)}$ and $T$ are Hermitian matrices, and $U_k$, $\alpha_k$ real numbers. 
 To simplify the notation, in the following we assume that the chemical potential term $\mu\hat{N}$, with $\hat{N} \equiv \sum_{x} \hat{c}^{\dagger}_{x} \hat{c}^{\phantom\dagger}_{x}$, has been adsorbed into the Hamiltonian $\hat{H}$.
 Using the Trotter  decomposition  with $L_\tau \Delta \tau = \beta$,  and a discrete version of the Hubbard-Stratonovich (HS) transformation, 
\begin{equation}
\label{HS_squares}
        e^{\Delta \tau  \lambda  \hat{A}^2 } = \frac{1}{4}
        \sum_{ l = \pm 1, \pm 2}  \gamma(l)
e^{ \sqrt{\Delta \tau \lambda }
       \eta(l)  \hat{A} }
                + {\cal O} (\Delta \tau ^4)\;,
\end{equation}
with $ \gamma(\pm 1)  = 1 + \sqrt{6}/3$, $\eta(\pm 1 ) = \pm \sqrt{\smash[b]{2 (3 - \sqrt{6} )}} $, and 
$  \gamma(\pm 2) = 1 - \sqrt{6}/3$, $\eta(\pm 2 ) = \pm \sqrt{\smash[b]{2 (3 + \sqrt{6} )}}$, 
 one can approximate  the imaginary  time propagator  $e^{-\beta \hat{H}}$ as 
 \begin{equation}
   \begin{split}
     e^{-\beta \hat{H} } =  &\sum_{ \left\{ l_{k,\tau} \right\} }   e^{S_0 \left\{ l_{k,\tau} \right\} }\prod_{\tau=1}^{L_{\tau}}   e^{-\Delta \tau \hat{T}}  \prod_{k}e^{  \sqrt{- \Delta \tau U_k } \eta(l_{k,\tau} )\hat{V}^{(k)} }\\
     & + i\Delta\tau \hat{R} + O(\Delta\tau^2).
   \end{split}
   \label{trotter}
\end{equation}
 Here $S_0 \left\{ l_{k,\tau} \right\}  =  \sum_{l_{k,\tau} } \ln \left( \gamma(l_{k,\tau} ) \right)  +    \sqrt{- \Delta \tau U_k } \eta(l_{k,\tau} ) \alpha_k $ . It is easy to show that the contribution of the anti-Hermitian operator $i\hat{R}$ to the expectation value of an Hermitian oberservable is purely imaginary, so that the discretization error $\propto \Delta\tau$ can be filtered out, leading to a Trotter error $\propto \Delta\tau^2$.
 The systematic  error involved in this discrete HS transformation  is of a higher order than the one encountered in the Trotter decomposition so that it can be regarded as good as exact.   
    At this point, one can integrate out the fermions  so as to obtain the grand-canonical  partition function: 
\begin{equation}
  Z_{\rm gc}  =  \Tr\left\{e^{-\beta \hat{H}} \right\}= \sum_{ \left\{ l_{k,\tau} \right\} }   e^{S_0 \left\{ l_{k,\tau} \right\} }    \det( 1 + U(l_{k,\tau} ) )
\end{equation} 
with 
\begin{equation}
	 U(l_{k,\tau} ) = \prod_{\tau=1}^{L_{\tau}}   e^{-\Delta \tau T}  \prod_{k}e^{  \sqrt{- \Delta \tau U_k } \eta(l_{k,\tau} )V^{(k)} }.
\end{equation}
Using the Leibniz formula for determinants, one can show that: 
\begin{equation}
  \begin{split}
    \det &( 1 +  U)    \\
&= 1 +    \sum_{N=1}^{N_s}\ \sum_{x_N > x_{N-1} > \cdots > x_1} \det
\begin{bmatrix} U_{x_1,x_1}  &  \dots & U_{x_1,x_N}  \\
                        \vdots &   \ddots & \vdots \\ 
                          U_{x_N,x_1}  & \cdots   & U_{x_N,x_N}   \end{bmatrix}\\
    &=  1 + \sum_x  U_{x,x}   + \sum_{x_2 > x_1}  \det 
 \begin{bmatrix} U_{x_1,x_1}  &  U_{x_1,x_2}  \\
                          U_{x_2,x_1}  & U_{x_2,x_2}   \end{bmatrix}  + \cdots
  \end{split}
  \end{equation}
Here $N_s$ corresponds to the number of single-particle states, and one can readily see that each term of the sum corresponds to the canonical trace of $N$ single-particle states. Thereby,  the canonical partition function  $Z_{\rm can}(N)$ is given by: 
\begin{equation}
	Z_{\rm can}(N) =   \frac{d^N}{dz^N} \sum_{ \left\{ l_{k,\tau} \right\} }\left.  e^{S_0 \left\{ l_{k,\tau} \right\} }     \det( 1 + z U(l_{k,\tau} ) ) \right|_{z=0}.
\end{equation}
A numerical implementation of the above equation reads: 
\begin{equation}
\begin{split}
	&Z_{\rm can}(N)\\
&=   \frac{1}{N_s}  \sum_{m=1}^{N_s} \sum_{ \left\{ l_{k,\tau} \right\} }   e^{S_0 \left\{ l_{k,\tau} \right\} }  e^{-i \phi_m N }  \det( 1 + e^{i \phi_m} U(l_{k,\tau} ) ),
\end{split}
\label{Z_can1.eq}
\end{equation}
where $\phi_m  = 2 \pi m/ N_s $. 
An equivalent way to show the above result is to note that  the total particle number $\hat{N}$ commutes with the Hamiltonian such that: 
\begin{equation}
\begin{split}
 Z_{\rm can}(N)   &=    {\rm Tr} \left[  \delta_{\hat{N}, N }  e^{-\beta \hat{H} }\right]\\ 
            &=\frac{1}{N_s} \sum_{m=1}^{N_s}   e^{-i \phi_m N }  {\rm Tr} \left[ e^{i \phi_m \hat{N}}e^{-\beta \hat{H} } \right].
\end{split}
\label{Zcan_alt}
\end{equation}
By applying a Trotter decomposition and HS transformation to the right-hand side of Eq.~(\ref{Zcan_alt}), one can reproduce   Eq.~(\ref{Z_can1.eq}).   Implementations of canonical simulations using the above results have been  proposed  in Refs.~\cite{Ormand94,Gilbreth15}.  In these approaches,  the discrete Fourier transformation is computed exactly  at each MC step.    For the method to be successful, the chemical potential has  to be chosen such that  the average particle number is peaked around the desired value. 

\subsection{Constraint of the particle-number fluctuations}
\label{sec:method:constraint}
Here we follow a  slightly different approach and modify the Hamiltonian as
\begin{equation}
  \begin{split}
    &\hat{H}(\lambda)  =  \hat{H} + \hat{H}_\lambda,\\
    &\hat{H}_\lambda \equiv \lambda  \left( \hat{N} - N_0 \right)^2,
  \end{split}
         \label{Ham_CE_QMC}
\end{equation}
 such that 
 \begin{equation}
 Z_{\rm can}(N_0)   = \lim_{\lambda \rightarrow \infty }  {\rm Tr} \left[   e^{-\beta \hat{H}(\lambda) } \right].
\label{Z_CE_QMC}
 \end{equation}
     As discussed above, in Eqs.~(\ref{Ham_CE_QMC}) and (\ref{Z_CE_QMC}) the Hamiltonian $\hat{H}$ implicitly depends on the chemical potential $\mu$, which needs to be tuned such that $\<\hat{N}\>=N_0$. In practice, this is done by computing $\<\hat{N}\>$ as a function of $\mu$, for a suitable interval in $\mu$, by means of auxiliary field QMC and then fixing $\mu$ such that the equation $\<\hat{N}\>=N_0$ is satisfied within the desired statistical accuracy; at half-filling one has exactly $\mu=0$.

     Since $\hat{H}$  conserves the particle number, one can foresee  rapid convergence in $\lambda$ because  particle-number sectors with $\hat{N} = N \neq N_0$ have a statistical weight suppressed  by a factor $ e^{-\lambda \beta \left( N - N_0 \right)^2 }$.    The latter also shows that the relevant parameter  for the convergence is $\beta \lambda$  rather than $\lambda $ itself.  The additional term  is a perfect square term which is  easily implemented within the ALF code \cite{ALF}.  Since  $ \left( \hat{N} -  N_0 \right)^{2}  $ effectively corresponds to a long-ranged interaction, one may   face the issue that the acceptance rate of a single  HS flip becomes excessively small on large lattices. To circumvent this problem we have  used the following decomposition:
\begin{equation}
        e^{-\beta \hat{H}}  =   \prod_{\tau = 1}^{L_{\tau}} \left[  e^{-\Delta \tau \hat{T}} e^{-\Delta \tau \hat{V}}
          \underbrace{e^{-\frac{\Delta \tau}{n_{\lambda}} \hat{H}_{\lambda} } \cdots e^{-\frac{\Delta \tau}{n_{\lambda}} \hat{H}_{\lambda} } }_{n_\lambda \text{-times } }\right].
        \label{trotter_nlambda}
\end{equation}
Thereby, we need $n_\lambda $ fields per time slice to impose the constraint. For each field, the coupling constant is effectively suppressed by a factor $n_{\lambda}$, thus allowing to control the acceptance of the QMC algorithm.

\begin{figure}
 \subfigure{\label{chi_c_1D_L}
 \includegraphics[width=8cm, height=6cm]{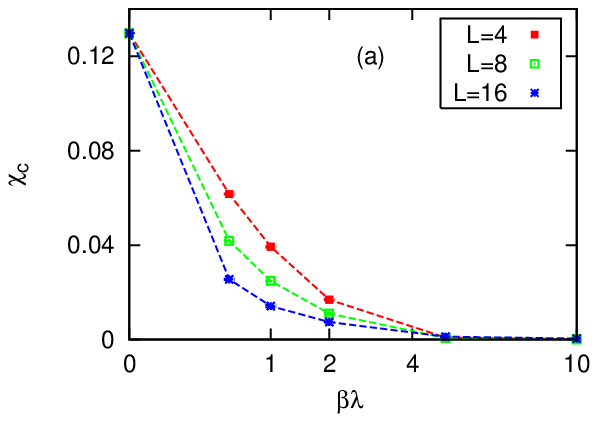}
 }
 \subfigure{
 \includegraphics[width=8cm, height=6cm]{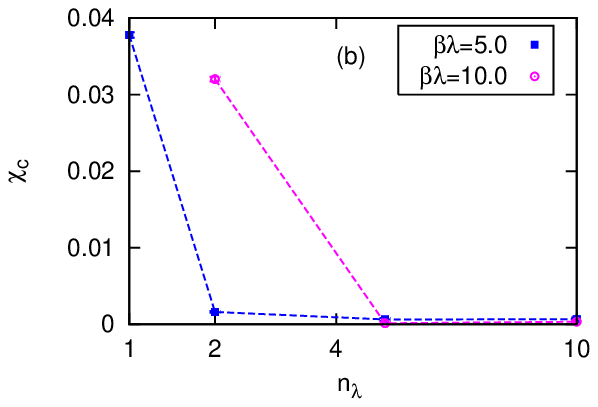}
 \label{chi_c_1D_L8}
 }
 \centering
 \caption{$\beta \lambda$ and $n_{\lambda}$ dependence of $\chi_c$ for the $1D$ Hubbard model at $U=4.0$ and $\beta=0.5$. 
   (a) $\chi_c$ as a function of $\beta \lambda$ for $L=4$, $8$, and $16$. For each $\beta \lambda$ we have taken the
   parameter $n_{\lambda}$ large enough as to effectively suppress the discretization error in the decomposition of the constraint.
   (b) $\chi_c$ as a function of $n_{\lambda}$ for $L=8$ and two values of $\lambda$. }
\label{chi_c_1D}
\end{figure}

In order to test the efficiency of our QMC method in the canonical ensemble, we computed the uniform intensive charge susceptibility $\chi_c$, defined as 
\begin{equation}
  \chi_c \equiv \frac{\beta}{V}\left(  \langle \hat{N}^2 \rangle - \langle \hat{N} \rangle^2  \right).
  \label{charge_chi}
\end{equation}
Note that compared with the extensive definition in Eq.~(\ref{charge_chi_extensive}), here the susceptibility is divided by the system volume $V$.
In Fig.~\ref{chi_c_1D} we show $\chi_c$ for the one-dimensional (1D) Hubbard model as a function of $\beta \lambda$ and $n_{\lambda}$.
As shown in Fig.~\ref{chi_c_1D}(a),
$\chi_c$ decays gradually from a finite value to zero on increasing $\beta\lambda$. 
The threshold in  $\lambda$ for which $\chi_c$ converges to zero corresponds to the canonical ensemble. 
A comparison of the results for lattice sizes $L=4$, $8$, and $16$ suggests that
the charge fluctuations are easier to  suppress for larger system sizes. The dependence of $\chi_c$ on $n_{\lambda}$ defined in Eq.~(\ref{trotter_nlambda})
is shown in Fig.~\ref{chi_c_1D}(b), which
illustrates the increased Trotter error for larger values of $\beta\lambda$.

\begin{figure}
 \includegraphics[width=8cm, height=6cm]{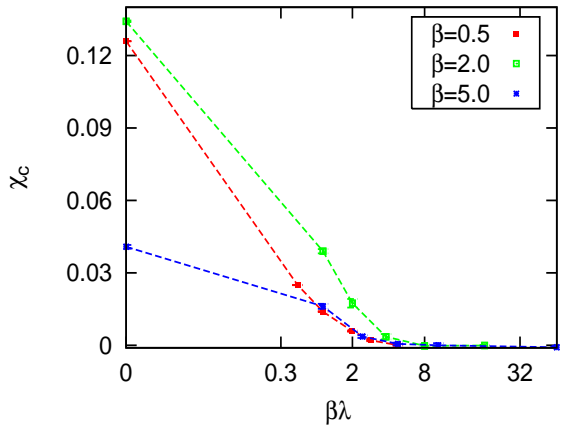}
 \centering
 \caption{ $\lambda$ dependence of $\chi_c$ for the $2D$ Hubbard model at $U=4.0$, $L=4$, and $\beta=0.5$, $2.0$, and $5.0$.}
\label{chi_c_2D_beta}
\end{figure}

Figure \ref{chi_c_2D_beta} shows the decay of charge susceptibility $\chi_c$ as a function of $\lambda$ 
in the two-dimensional (2D) Hubbard model, for $U=4.0$, $L=4$ and
several inverse temperatures $\beta=0.5$, $2.0$, and $5.0$.
Inspection of Fig.~\ref{chi_c_2D_beta} reveals that in the
 grand-canonical ensemble the $\beta=2.0$ case exhibits charge fluctuations larger than the $\beta=0.5$ case, thereby requiring a larger value of $\beta\lambda$
 to realize the canonical ensemble.

\section{Finite-size corrections in the canonical ensemble: exact results}
\label{sec:fscanonical}
In this section, by exploiting the relation between the canonical and the grand-canonical free energy, we determine the leading finite-size correction of the free energy in the canonical ensemble and relate it to a susceptibility. To be concrete, we consider a quantum model on a lattice, where in the canonical ensemble the number of particles is fixed, and we prove that on a finite volume $V$
\begin{equation}
\begin{split}
  F_{\rm can}&(n_0,V) - F_{\rm gc} (V)\\
&= \frac{1}{2V}\ln\left(2\pi V\right) + \frac{1}{2V}\ln\left(\frac{\chi_c}{\beta}\right) + O\left(\frac{1}{V^2}\right),
\end{split}
\label{F_can_diff_gc}
\end{equation}
where $F_{\rm can}(n_0,V)$ and $F_{\rm gc} (V)$ are the free energies per volume $V$ and in units of $k_BT$ in the canonical and grand-canonical ensembles, respectively, and $\chi_c$ is the charge susceptibility (in the grand-canonical ensemble), defined in Eq.~(\ref{charge_chi}); the filling fraction $n_0$ in $F_{\rm can}(n_0,V)$ is fixed to the corresponding expectation value in the grand-canonical ensemble. Equation (\ref{F_can_diff_gc}) provides the leading additional contribution to the free energy density due to the particle-number constraint. As discussed towards the end of this section, Eq.~(\ref{F_can_diff_gc}) allows also to determine the leading finite-size correction of observables in the canonical ensemble if, as expected, finite-size corrections in the grand-canonical ensemble decay faster than $1/V$.

In order to prove Eq.~(\ref{F_can_diff_gc}), we observe that the free energy density $F_{\rm can}(n,V)$ can be related to a path-integral formulation of the canonical partition function as
\begin{equation}
e^{-VF_{\rm can}(n,V)} = \int [{\cal D}\Psi] e^{-S[\Psi]}\delta\left(n,\frac{1}{V}\hat{N}(\Psi)\right),
\label{Fcan}
\end{equation}
where $\Psi$ indicates collectively the fields entering in the path integral, $S[\Psi]$ is the action of the model, $\hat{N}(\Psi)$ is the expression of the total number operator $\hat{N}$ in terms of the fields $\Psi$, and $n$ is the intensive filling fraction, which is fixed in the canonical ensemble. In Eq.~(\ref{Fcan}), $S[\Psi]$, as well as $F_{\rm can}(n,V)$, additionally depend on the temperature and coupling constants, inessential for the present discussion.
On a lattice, $\hat{N}$ is the sum of single-site and single-species number operators $\hat{N}(x)$, $\hat{N}=\sum_x \hat{N}(x)$, therefore $n$ can only take discrete values, separated by an interval of $1/V$. By summing over the allowed values of $n$, we obtain the grand-canonical free energy density $F_{\rm gc}(V)$
\begin{equation}
e^{-VF_{\rm gc}(V)} = \sum_{n=n_{\rm min}}^{n_{\rm max}} e^{-VF_{\rm can}(n,V)},
\label{F_gc_discrete}
\end{equation}
where, as before, we have ignored the dependence of $F_{\rm gc}(V)$ on the various coupling constants, and $n_{\rm min}$, $n_{\rm max}$ indicate the minimum and maximum number of particles per volume that the model can host; usually $n_{\rm min}=0$, while $n_{\rm max}$ depend on the number and type of particle species. For $V\rightarrow\infty$ the sum in Eq.~(\ref{F_gc_discrete}) can be approximated by the Euler-Maclaurin formula as
\begin{equation}
\begin{split}
&e^{-VF_{\rm gc}(V)} = V\Bigg[\int_{n_{\rm min}}^{n_{\rm max}} dn\ e^{-VF_{\rm can}(n,V)}\\
&+ \frac{e^{-VF_{\rm can}(n_{\rm min},V)}+e^{-VF_{\rm can}(n_{\rm max},V)}}{2V}+O\left(\frac{e^{-cV}}{V}\right)\Bigg],
\end{split}
\label{F_gc_continous}
\end{equation}
where the next-to-leading term in the Euler-Maclaurin formula is $\propto (1/V^2)\partial (e^{-VF_{\rm can}})/\partial n$ computed at the end points, hence it is of order $e^{-cV}/V$.
In the limit $V\rightarrow\infty$, the integral on the right-hand side of Eq.~(\ref{F_gc_continous}) is dominated by the minimum $n_0$ of $F_{\rm can}(n,V)$. If $n_0$ is an interior point\footnote{The case of multiple saddle points, or a saddle point at an end point requires a separate analysis.} of the integration interval $[n_{\rm min},n_{\rm max}]$, by using the saddle-point method we obtain
\begin{equation}
\begin{split}
&e^{-VF_{\rm gc}(V)}/V = e^{-VF_{\rm can}(n_0,V)}\cdot\\
&\left[\frac{2\pi}{V(\partial^2F_{\rm can}/\partial n^2)(n_0,V)}\right]^{1/2}\left[1+O\left(\frac{1}{V}\right)\right]\\
&+ \frac{e^{-VF_{\rm can}(n_{\rm min},V)}+e^{-VF_{\rm can}(n_{\rm max},V)}}{2V}+O\left(\frac{e^{-cV}}{V}\right),
\end{split}
\label{F_gc_continous_saddle}
\end{equation}
where the factor $1+O(1/V)$ represents the next-to-leading term in the saddle-point expansion.
The second term on the right-hand side of Eq.~(\ref{F_gc_continous}) is depressed by a factor $\propto \exp\{-V[F_{\rm can}(n_{\rm min},V)-F_{\rm can}(n_0,V)]\}/V^{1/2} + \exp\{-V[F_{\rm can}(n_{\rm max},V)-F_{\rm can}(n_0,V)]\}/V^{1/2}$ with respect to the first term, therefore, since $n_0$ is the minimum of $F_{\rm can}(n,V)$, it is subleading with respect to the first factor. Moreover, the convergence of the integral in Eq.~(\ref{F_gc_continous}) requires the last term on the right-hand side of Eq.~(\ref{F_gc_continous_saddle}) to be subleading with respect to the first factor. Thus, by factorizing the first term on the right-hand side of Eq.~(\ref{F_gc_continous_saddle}) and taking the logarithm, the last two terms give a contribution of order $\ln(1+\exp\{-cV\}/V^{1/2})\sim \exp\{-cV\}/V^{1/2}$, which is negligible with respect to the correction of order $1/V$ originating from the next-to-leading term of the saddle-point expansion. 
On taking the logarithm on both sides of Eq.~(\ref{F_gc_continous_saddle}) we find
\begin{equation}
\begin{split}
F_{\rm gc}(V) &= F_{\rm can}(n_0,V) - \frac{1}{V}\ln V - \frac{1}{2V}\ln\left(\frac{2\pi}{V}\right) \\
&+ \frac{1}{2V}\ln\left[\frac{\partial^2F_{\rm can}}{\partial n^2}(n_0,V)\right] + O\left(\frac{1}{V^2}\right),
\end{split}
\label{F_gc_vs_can}
\end{equation}
where subleading exponential corrections have been neglected. The second and third terms $\propto \ln V$ on the right-hand side of Eq.~(\ref{F_gc_vs_can}) represent an entropic contribution which is due to the larger configurational space of the grand-canonical ensemble as compared to the canonical one. In particular, the first constant originates from the discretization of the allowed values of $n$ [see the discussion after Eq.~(\ref{Fcan})] and is absent in continuous models.
The saddle-point position $n_0$ appearing in the previous equations corresponds precisely to the grand-canonical expectation value of $\<\hat{N}/V\>_{\rm gc}$. This is because, using Eq.~(\ref{Fcan}) and Eq.~(\ref{F_gc_discrete}), one can write $\<\hat{N}/V\>_{\rm gc}e^{-VF_{\rm gc}(V)} = \sum_{n=n_{\rm min}}^{n_{\rm max}} ne^{-VF_{\rm can}(n,V)}$. Along the same line of reasoning as above, one finds that, as expected also from thermodynamic considerations, $\lim_{V\rightarrow\infty}\<\hat{N}/V\>_{\rm gc} = n_0$. Thus, the quantity $F_{\rm can}(n_0,V)$ on the right-hand side of Eq.~(\ref{F_gc_vs_can}) is precisely the free energy density with a particle number fixed to its expectation value in the grand-canonical ensemble, i.e., the thermodynamic quantity which is meaningful to compare with the grand-canonical free energy density. The fluctuation of the particle number, which determines the charge susceptibility $\chi_c$ defined in Eq.~(\ref{charge_chi}), can be related to the finite-size correction on the right-hand side of Eq.~(\ref{F_gc_vs_can}). By using Eq.~(\ref{Fcan}), Eq.~(\ref{F_gc_discrete}), and the definition of Eq.~(\ref{charge_chi}), one obtains
\begin{equation}
\chi_c=\frac{\sum\limits_{n=n_{\rm min}}^{n_{\rm max}}\frac{\beta}{V}(nV-n_0V)^2 e^{-VF_{\rm can}(n,V)}}{\sum\limits_{n=n_{\rm min}}^{n_{\rm max}} e^{-VF_{\rm can}(n,V)}}.
\label{n_fluctuation}
\end{equation}
The right-hand side of Eq.~(\ref{n_fluctuation}) can be evaluated for $V\rightarrow\infty$ using a saddle-point expansion as above, resulting in
\begin{equation}
\chi_c \underset{V\rightarrow\infty}{=} \frac{\beta}{\left(\partial^2F_{\rm can}/\partial n^2\right)(n_0,V)}.
\label{n_fluctuation_result}
\end{equation}
Finally, inserting Eq.~(\ref{n_fluctuation_result}) into Eq.~(\ref{F_gc_vs_can}), we obtain Eq.~(\ref{F_can_diff_gc}).

If finite-size corrections of $F_{\rm gc}(V)$ decay faster than $1/V$ (indeed, as discussed in Sec.~\ref{sec:intro}, we expect exponentially decaying finite-size corrections), we can replace $F_{\rm gc}(V)$ on the left-hand side of Eq.~(\ref{F_can_diff_gc}) with its ensemble-independent thermodynamic limit $F(V=\infty) = F_{\rm gc}(V=\infty) = F_{\rm can}(n_0,V=\infty)$, such that the leading finite-size corrections in $F_{\rm can}(n_0,V)$ are
\begin{equation}
\begin{split}
F_{\rm can}&(n_0,V) - F(V=\infty)\\
&= \frac{1}{2V}\ln\left(2\pi V\right) + \frac{1}{2V}\ln\left(\frac{\chi_c}{\beta}\right) + O\left(\frac{1}{V^2}\right).
\end{split}
\label{F_can_finitesize}
\end{equation}
From Eq.~(\ref{F_can_finitesize}) we can, e.g., determine the leading finite-size correction of the energy density in the canonical ensemble by taking the derivative with respect to $\beta$:
\begin{equation}
E_{\rm can}(V) - E(V=\infty) =  \frac{\partial\left(\chi_c/\beta\right)/\partial\beta}{2V\left(\chi_c/\beta\right)}.
\label{E_can_finitesize}
\end{equation}
It is useful to remark that the charge susceptibility $\chi_c$ appearing in Eqs.~(\ref{F_can_diff_gc}) and (\ref{n_fluctuation})-(\ref{E_can_finitesize}) is computed in the grand-canonical ensemble. Since $\chi_c$ has a finite thermodynamic limit and exponentially decaying finite-size corrections, it does not give rise to a further algebraic volume dependence.
  Equation ~(\ref{E_can_finitesize}) can be generalized to other local observables and correlations thereof. To this end, one can supplement the action of the model with an external source term
  \begin{equation}
    S[\Psi] \rightarrow S[\Psi] - h\sum_{x}\int d\tau O(x,\tau),
    \label{action_with_h}
  \end{equation}
  where the sum extends to the lattice sites and $O(x,\tau)$ is a local observable, to be expressed in terms of the fields $\Psi$ entering in the path integral of Eq.~(\ref{Fcan}). Such an addition corresponds to the insertion of external lines in the Feynman diagram expansion, and hence one expects, in line with the analysis of Ref.~\cite{Neuberger-89}, that in the presence of a finite mass gap correlations including $O(x,\tau)$ are characterized by exponentially decaying finite-size corrections. Under the substitution of Eq.~(\ref{action_with_h}), the charge susceptibility $\chi_c$ entering in Eq.~(\ref{F_can_diff_gc}) and Eq.~(\ref{F_can_finitesize}) acquires a dependence on the external field $h$. Differentations of the free energy density with respect to $h$ provide the analogous of Eq.~(\ref{E_can_finitesize}) for the finite-size corrections of the volume-average and susceptibility of $O$:
  \begin{equation}
    \begin{split}
      &O \equiv \frac{1}{\beta V} \sum_{x}\int d\tau \<O(x,\tau)\>_{h=0},\\
    &O_{\rm can}(V) - O(V=\infty) =  -\frac{\partial\chi_c(h)/\partial h|_{h=0}}{2\beta V\chi_c(h=0)},
    \end{split}
    \label{O_can_finitesize}
  \end{equation}
  \begin{equation}
    \begin{split}
      &\chi_O \equiv \frac{1}{\beta V} \sum_{x, x'}\int d\tau d\tau' \Big[\<O(x,\tau)O(x',\tau')\>_{h=0}\\
      &\qquad\qquad\qquad\qquad\qquad - \<O(x,\tau)\>\<O(x',\tau')\>_{h=0}\Big],\\
      &\chi_{O,{\rm can}}(V) - \chi_O(V=\infty)\\
      &= \frac{\left(\partial\chi_c(h)/\partial h|_{h=0}\right)^2-\chi_c(h=0)\partial^2\chi_c(h)/\partial h^2|_{h=0}}{2\beta V\chi_c(h=0)^2},
    \end{split}
    \label{chiO_can_finitesize}
  \end{equation}
  where we emphasize that the expectation values of $O(x,\tau)$ are computed in absence of the external field $h$.
  In particular, in a spinful model Eq.~(\ref{chiO_can_finitesize}) implies a leading finite-size correction $\propto 1/V$ of the spin susceptibility in the canonical ensemble. 
  We remark that the derivatives of $\chi_c(h)$ appearing in Eqs.~(\ref{O_can_finitesize}) and (\ref{chiO_can_finitesize}) can be in principle directly computed by sampling a suitable observable, thus avoiding a numerical differentiation.
  
  The results of Eqs.~(\ref{F_can_diff_gc}), (\ref{F_can_finitesize}), (\ref{E_can_finitesize}), (\ref{O_can_finitesize}), and (\ref{chiO_can_finitesize}) can be easily generalized to other correlations by considering a considering a space- and imaginary time-dependent source $h(x,\tau)$ in Eq.~(\ref{action_with_h}), or to other types of constrained models, along the same line of reasoning.

\section{Fermionic simulations  in the  canonical ensemble}
\label{sec:qmc}
We performed QMC simulation of the $SU(2)$ Hubbard model in both the grand-canonical and canonical ensemble. The Hamiltonian of the Hubbard model is defined as:
\begin{align}
 \hat{H} =  &-t \sum_{< \bm i, \bm j>, \sigma} \hat{c}^{\dagger}_{\bm i, \sigma}  \hat{c}^{\phantom\dagger}_{\bm j, \sigma}
 + U \sum_{\bm i} \left(\hat{n}_{\bm i, \uparrow} -\frac{1}{2} \right) \left(\hat{n}_{\bm i, \downarrow} -\frac{1}{2} \right) \nonumber\\
 &-\mu\sum_{\bm i}\left(\hat{n}_{\bm i, \uparrow}+\hat{n}_{\bm i, \downarrow}\right),
\label{Hubbard_Hamiltonian}
\end{align}
where $\hat{n}_{\bm i, \sigma}\equiv \hat{c}^{\dagger}_{\bm i, \sigma}  \hat{c}_{\bm i, \sigma }$. The canonical ensemble is realized by adding the constraint given in Eq.~(\ref{Ham_CE_QMC}). For such a modified Hamiltonian, the total number of  particles 
converges quickly to $N_0$ on increasing $\beta \lambda$.

Here we simulated both ensembles on a 1D lattice, as well as on the 2D square lattice at finite temperature, both of which are known to be disordered.
We mainly considered the models at half filling ($N_0=N_s/2$, with $N_s=2L^d$)
with zero chemical potential $\mu=0$ and carried out some test calculations  for the two-dimensional  doped Hubbard model. In all simulations we fixed $t=1$ and $U=4.0$.
Our basic MC observables are as follows:
\begin{enumerate}
\item Energy density\footnote{Up to an inessential, filling-dependent, additive constant, $E$ corresponds to the energy part on the right-hand side of Eq~(\ref{Hubbard_Hamiltonian}).}:
\begin{equation}
E=\frac{1}{L^d} \Bigg\langle -t \sum_{< \bm i, \bm j>, \sigma } \hat{c}^{\dagger}_{\bm i, \sigma}  \hat{c}^{\phantom\dagger}_{\bm j, \sigma }
+ U \sum_{\bm i} \hat{n}_{\bm i, \uparrow} \hat{n}_{\bm i, \downarrow} \Bigg\rangle
\end{equation}

\item Uniform spin susceptibility:
 \begin{equation}
  \chi_s = \frac{\beta}{L^d} \sum_{\bm i, \bm j} \<\hat{S}_{\bm i} \hat{S}_{\bm j}\>
 \end{equation}
\end{enumerate}

\begin{figure}
\includegraphics[width=8cm, height=6cm]{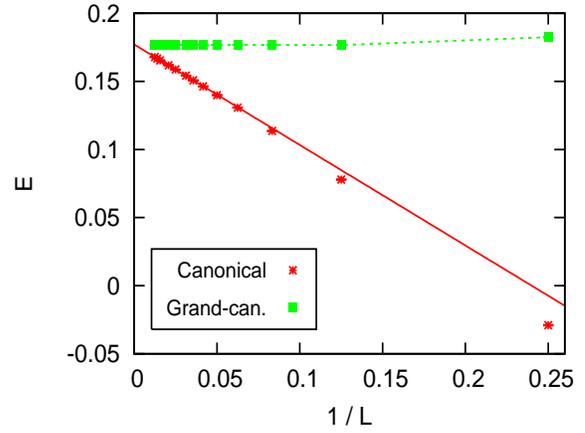}
\centering
\caption{Finite-size data of the energy density $E$ for the 1D Hubbard model in the grand-canonical and canonical ensembles, at $\beta=0.5$ and half-filling. The red line is a linear fit of
the canonical ensemble data to $E_{\rm can}(L)=E(L \rightarrow \infty)+a/L$, with $E(L \rightarrow \infty)=0.1771(2)$ and $a=-0.738(4)$, 
where the minimum lattice size taken into account is $L_{\rm min}=16$;
the dashed green line linking the grand-canonical data is a guide to the eye.}
\label{Energy_1D}
\end{figure}

\begin{figure}[b]
 \includegraphics[width=8cm, height=6cm]{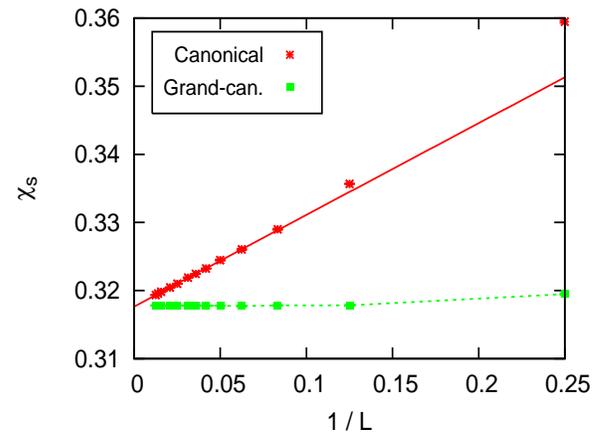}
 \centering
 \caption{Same as Fig.~\ref{Energy_1D} for the spin susceptibility $\chi_s$.  The red line is the linear fit of the
canonical ensemble data to $\chi_{s, {\rm can}}(L)= \chi_s (L \rightarrow \infty) + a/L$, with $\chi_s(L \rightarrow \infty)= 0.3177(1)$ and $a=0.135(1)$, where the minimum lattice size taken into account is $L_{\rm min}=12$.}
\label{Spinsus_1D}
\end{figure}

\subsection{1D model}
The QMC simulations of the one-dimensional Hubbard model are performed in both the grand-canonical and canonical ensembles at inverse temperature $\beta=0.5$,
system sizes $L=4$, $8$, $12$, $16$, $20$, $24$, $28$, $32$, $40$, $48$, $64$, $72$, $80$, and at half-filling.
 A comparison of the size effect for the energy density $E(L)$ and for the uniform spin susceptibility $\chi_s(L)$
in the two ensembles is shown in Fig.~\ref{Energy_1D} and Fig.~\ref{Spinsus_1D}, respectively. 
We observe that in the grand-canonical ensemble both
$E$ and $\chi_s$ converge quickly to the thermodynamic limit for
 small system sizes.
This indicates a small correlation length $\xi$ at this temperature.

On the other hand, except for the smallest system sizes, in the canonical ensemble
both observables exhibit a linear-like behavior as a function of $1/L$.
A fit of energy density in the canonical ensemble $E_{\rm can}(L)$ to
$E_{\rm can}(L)= E(L \rightarrow \infty) + a L^{-1} $
exhibits a good $\chi^2/\text{DOF}$ ($\text{DOF}$ denotes the number of degrees of freedom),
when the data for the small sizes are discarded;
the extrapolated value $E(L \rightarrow \infty)$ matches the grand-canonical result.
Similar considerations hold for a fit of the spin susceptibility in the canonical ensemble $\chi_{s, {\rm can}}(L)$ to $\chi_{s,{\rm can}}(L)= \chi_s (L \rightarrow \infty) + a L^{-1}$.

 Moreover, a fit of $E_{\rm can}(L)$ to $E(L\rightarrow \infty) + a L^{-d}$, leaving $d$ as a free parameter,
 gives $d=1.05(2)$ when the smallest lattice size taken into account for the fit is
 $L_{\rm min}=16$. An equivalent fit for $\chi_s(L)$ gives $d=1.04(2)$, when $L_{\rm min}=12$. 
 This confirms that finite-size corrections of observables in the canonical ensemble are $\propto 1/L$.

\begin{figure}
 \includegraphics[width=8cm, height=6cm]{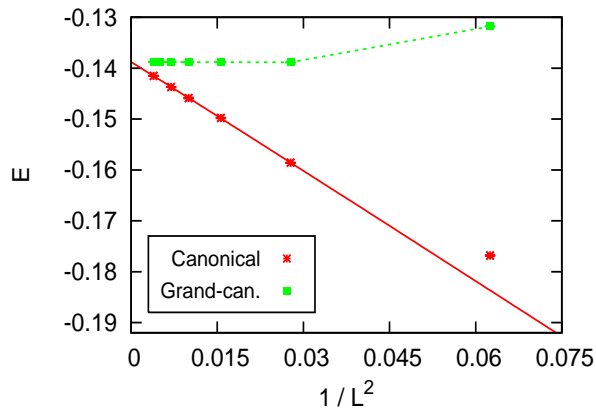}
 \centering
 \caption{ Same as Fig.~\ref{Energy_1D}  for the 2D Hubbard model. 
 The red line is a linear fit of the canonical ensemble data to $E_{\rm can}(L)=E(L \rightarrow \infty)+a/L^2$, with $E(L \rightarrow \infty)=-0.1387(1)$ and $a=-0.714(2)$
 , where the minimum lattice size taken into account is $L_{\rm min}=6$. }
\label{Energy_2Db05}
\end{figure}

\begin{figure}[b]
 \includegraphics[width=8cm, height=6cm]{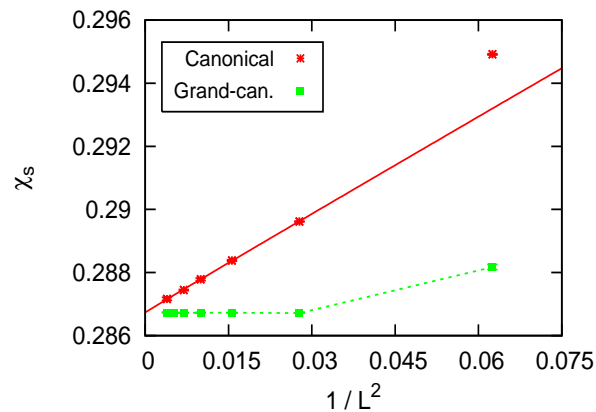}
 \centering
  \caption{Same as Fig.~\ref{Energy_2Db05} for the spin susceptibility $\chi_s$. 
 The red line is a linear fit of the canonical ensemble data to $\chi_{s, {\rm can}}(L)= \chi_s (L \rightarrow \infty) + a/L^2$, with $\chi_s(L \rightarrow \infty)= 0.2867(1)$ and $a=0.104(1)$
 , where the minimum lattice size taken into account is $L_{\rm min}=6$. }
\label{Spinsus_2Db05}
\end{figure}

\subsection{2D model}

We simulated the Hubbard model on the square lattice for both ensembles at $\beta=0.5$ and $\beta=2.0$, lattice sizes $L=4$, $6$, $8$, $10$, $12$, $14$, $16$, and at half-filling.

Figure \ref{Energy_2Db05} and Fig.~\ref{Spinsus_2Db05}
show the size behavior of $E$ and $\chi_s$ for the two ensembles at $\beta=0.5$.
The observed tiny size dependence of the observables in the grand-canonical ensemble suggests that the correlation length $\xi$
is smaller than the minimum lattice size $L=4$.
On the other hand, in the canonical ensemble
the energy density $E$ and the spin susceptibility $\chi_s$ show a linear-like behavior as function of $1/L^2$.

For a more quantitative check of the finite-size
correction in the canonical ensemble, we fitted $E_{\rm can}(L)$ to
$E_{\rm can}(L)= E(L\rightarrow\infty) + aL^{-1} +bL^{-2} + cL^{-3} $ 
and $\chi_{s, {\rm can}}(L)$ to an equivalent Ansatz, leaving $a,b$ and $c$ as free parameters. 
 Fit results for both observables show a good $\chi^2/\text{DOF}$ when $L_{\rm min}=6$, and the coefficient $a$ vanishes within error bars, whereas $b$ acquires a finite value.
 On the other hand, a fit of $E_{\rm can}(L)$ to $E_{\rm can}(L)=E(L\rightarrow \infty) + bL^{-d}$, leaving $b$ and $d$ as free parameters, and of $\chi_s(L)$ to an equivalent Ansatz,
gives $d=2.05(3)$ and $d=1.9(1)$ for $E$ and $\chi_s$, respectively, when $L_{\rm min}=6$.  
In line with the discussions of Sec.~\ref{sec:fscanonical},
these fit results confirm that the leading finite-size correction in the 
canonical ensemble is $\propto 1/L^2$.

\begin{figure}[b]
 \centering
 \includegraphics[width=8cm, height=6cm]{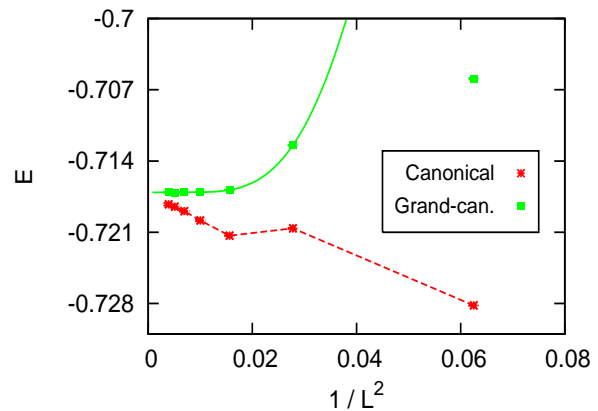}
\caption{Same as Fig.~\ref{Energy_2Db05} for $\beta=2.0$. The green line is the exponential fit of
the grand-canonical ensemble data with minimum size $L_{\rm min}=6$ and parameters $E(L\rightarrow\infty)=-0.717075$, $b=36$ and $c=0.67$ (see main text). }
\label{Energy_2Db2}
\end{figure}

We also simulated the 2D Hubbard model at a lower temperature $\beta=2.0$. A corresponding comparison of the finite-size energy density for the grand-canonical and canonical ensemble is shown in Fig.~\ref{Energy_2Db2}.
Generically, finite-size corrections in the canonical ensemble are expected to be temperature dependent.
On the other hand, the exponential correction characterized by the correlation length in the grand-canonical ensemble may start to be relevant at a lower temperature, because of an increased correlation length.

The data shown in Fig.~\ref{Energy_2Db2} exhibit a visible decay of the energy density in the grand-canonical ensemble $E_{\rm gc}$, on increasing the system size. 
 As a guide to the eye, we fitted $E_{\rm gc}$ to $E_{\rm gc}(L)= E(L\rightarrow \infty)+ b\cdot e^{-L/c}$.
 The finite-size values of $E$ in the canonical ensemble show a nonmonotonic behavior between $L=4$ and $6$, which might be due
 to a combination of various sources of finite-size corrections, such as the one $\propto 1/V$ originating from the particle-number constraint, the one related to the correlation length, and the residual correction term due to the regular part of the free energy.
 Nevertheless, a finite-size dependence $\propto 1/L^2$ can be clearly observed in Fig.~\ref{Energy_2Db2} for $L>6$, with a smaller slope compared to the $\beta=0.5$ case (compare with Fig.~\ref{Energy_2Db05}).

\begin{figure}
 \includegraphics[width=8cm, height=6cm]{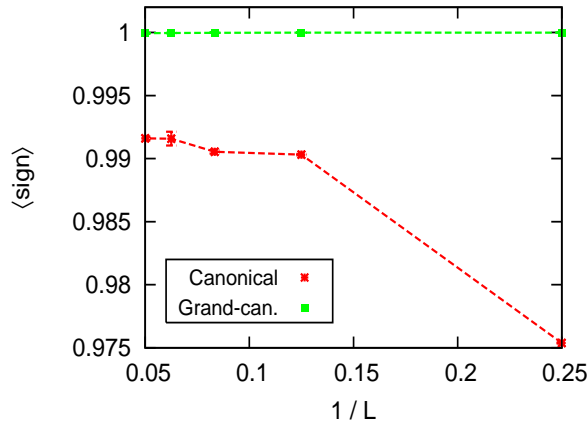}
 \centering
\caption{Finite-size data of MC average sign $\<\text{sign}\>$ for the 2D Hubbard model at $\beta=2.0$ in the grand-canonical and canonical ensemble, with a $\frac{3}{8}$-filling fraction and for lattice sizes up to $L=20$.}
\label{Sign_2Db2_doping}
\end{figure}

\begin{figure}[b]
 \includegraphics[width=8cm, height=6cm]{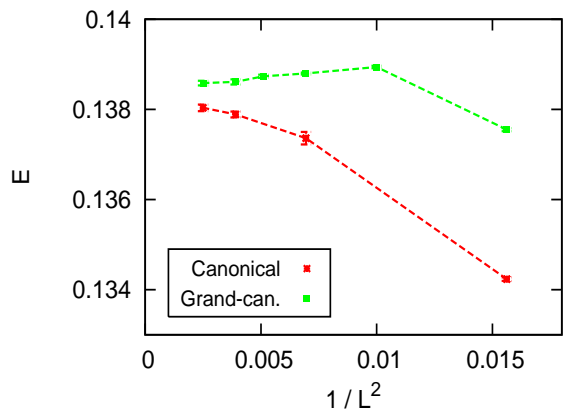}
 \centering
\caption{Finite-size data of energy density $E$ for the same case as in Fig.~\ref{Sign_2Db2_doping}.}
\label{Energy_2Db2_doping}
\end{figure}

We note that the auxiliary field QMC for the grand-canonical ensemble has a mild sign-problem under doping the system away from half filling, provided that the temperature is high enough.
Here we also tested the efficiency of the canonical ensemble QMC method under doping.
To this end, for every lattice size we tuned the chemical potential $\mu$ such that the expectation value of the number of particles in the grand-canonical ensemble
matches the desired number $N_0$ of particles in the canonical ensemble.
Subsequently, the canonical ensemble is realized by introducing a Lagrange multiplier, as discussed in Sec.~\ref{sec:method:constraint}.
In order to test the sign-problem, we also calculated the average sign during the MC simulation:

\begin{equation}\label{def_sign}
 \langle \text{sign} \rangle = \frac{\sum_C Re[e^{-S(C)}]}{\sum_C |Re[e^{-S(C)}]|}
\end{equation}
where $S(C)$ is the action for the MC configuration $C$, so that the corresponding statistical weight is $\propto e^{-S(C)}$.
The sign is not necessarily 
a real number (when the MC is sign-problem free, $S(C)$ is real and $\langle \text{sign} \rangle =1$).
On the other hand, the expectation value of observables can be computed via a reweighting scheme only when $\langle \text{sign} \rangle $ 
is not too small.
Figure \ref{Sign_2Db2_doping} shows the average sign during MC of a doped 2D Hubbard model at $\frac{3}{8}$ filling, which in the grand-canonical ensemble system does not exhibit a significant sign-problem at an inverse temperature of $\beta=2.0$.
Quite remarkably, when the number of particles is fixed in the canonical ensemble,
the code still exhibits an average sign higher than $0.97$, for system sizes up to $L=20$. 
This confirms the feasibility of our QMC method for the canonical ensemble, even under doping.
Figure \ref{Energy_2Db2_doping} shows the finite-size behavior of energy density of the two ensembles at $\frac{3}{8}$ filling. Similar to the results at half-filling, and in line with the analysis of Sec.~\ref{sec:fscanonical}, the energy density in the grand-canonical ensemble exhibits very small finite-size corrections when $L\ge 10$, whereas in the canonical ensemble we observe a finite-size correction approximately linear in $1/L^2$ for $L\ge 12$.

\section{Summary}
\label{sec:conclusions}
In this paper we have introduced a method to simulate fermionic models in the canonical ensemble. It consists in an auxiliary field QMC simulation, where the Hamiltonian is supplemented by an additional Lagrange multiplier, which constraints the particle number. The method can implemented using the ALF package for fermionic simulations \cite{ALF}.
In general, we find that canonical simulations are more computationally demanding than the corresponding ones in the grand-canonical ensemble.
 Although in the presence of short-ranged interactions the grand-canonical and the canonical ensemble are equivalent in the thermodynamic limit, their approach to the infinite-volume limit is distinctively different. In the canonical ensemble the observables are generically found to display a finite-size correction which is proportional to the inverse volume. In Sec.~\ref{sec:fscanonical} we prove an exact formula for the leading finite-size correction of the free energy density, the energy density, and other observables. Such a correction is controlled by the charge susceptibility and is found to be proportional to the inverse volume. This result is further substantiated by an exact calculation for the one-dimensional Ising model reported in the Appendix. Our numerical simulations of the Hubbard model reported in Sec.~\ref{sec:qmc} confirm the presence of finite-size corrections proportional to the inverse volume in the canonical ensemble. In line with previous theoretical results, in the presence of a finite correlation length and for periodic boundary conditions, observables computed in the grand-canonical ensemble display a faster approach to the thermodynamic limit, such that the leading finite-size correction is exponential in the ratio of the linear size over the correlation length.

\noindent 
{\bf Note  added}:   After completing  this paper we became aware of related research presented in Ref.~\cite{GGD-17}, which investigates the effect of a constraint within statistical field theory.

\acknowledgments
FFA acknowledges useful conversations with A. Sandvik. FPT acknowledges useful communications with M. Gross. This work was supported by the German Research Foundation (DFG) through Grant No. SFB 1170 ToCoTronics and Grant No. FOR 1807. We acknowledge the computing time granted by the John von Neumann Institute for Computing (NIC) and provided on the supercomputer JURECA \cite{Jureca16} at the J\"ulich Supercomputing Centre.

\appendix
\renewcommand{\theequation}{A\arabic{equation}}
\addcontentsline{toc}{section}{Appendix}
\setcounter{section}{0}
\setcounter{equation}{0}
\section*{Appendix: One-dimensional Ising model in the canonical ensemble: Exact results}
\label{sec:ising}
In this Appendix we compute the leading finite-size correction of the free energy of the one-dimensional Ising model in the canonical ensemble.
Although we mainly consider the antiferromagnetic Ising model, the results are also valid for the ferromagnetic model. Employing periodic boundary conditions, the Hamiltonian is
\begin{equation}
{\cal H} = J\sum_{i=1}^{L} S_iS_{i+1},\quad S_i=\pm 1
\end{equation}
where $L$ is the number of sites and $S_{L+1}\equiv S_1$. The partition function $Z_{\rm can}$ in the canonical ensemble with fixed magnetization $M=0$ is given by
\begin{equation}
  Z_{\rm can} = \sum_{\{S_k=\pm 1\}}\exp\left\{-K\sum_{i=1}^LS_iS_{i+1}\right\}\delta\left(\sum_{i=1}^L S_i,0\right),
  \label{ZIsing1D_def}
\end{equation}
where we have defined $K\equiv \beta J$.

The constraint $M=0$ can be expressed by using an integral representation of the Kronecker delta function $\delta$ appearing in Eq.~(\ref{ZIsing1D_def}), such that
\begin{equation}
\begin{split}
  &Z_{\rm can}\\
&= \frac{1}{2\pi}\sum_{\{S_k=\pm 1\}}\int_0^{2\pi} d\mu\exp\left\{-K\sum_{i=1}^LS_iS_{i+1} + i\mu\sum_{i=1}^L S_i\right\}.
\end{split}
  \label{ZIsing_integral}
\end{equation}
Inspecting Eq.~(\ref{ZIsing_integral}), we observe that $Z_{\rm can}$ is obtained as the integral over $\mu$ of the partition function for a one-dimensional Ising model in an external imaginary field $i\mu$. The trace over the configuration space can be computed using standard transfer-matrix techniques, yielding
\begin{equation}
  Z_{\rm can} = \frac{1}{2\pi}\int_0^{2\pi} d\mu\left[\lambda_+(\mu)^L+\lambda_-(\mu)^L\right],
  \label{ZIsing_eigenvalues}
\end{equation}
where the eigenvalues of the transfer matrix are
\begin{equation}
  \lambda_\pm(\mu) = e^{-K}\left(\cos(\mu)\pm \sqrt{e^{4K}-\sin(\mu)^2}\right),
  \label{Ising_lambda}
\end{equation}
and $\lambda_\pm(\mu)$ depends implicitly also on $K$. By noting that $\lambda_\pm(\mu+\pi) = -\lambda_{\mp}(\mu)$, Eq.~(\ref{ZIsing_eigenvalues}) can be cast in the form
\begin{equation}
  \begin{split}
    Z_{\rm can} &= \frac{1}{2\pi}\int_0^{\pi} d\mu\left[\lambda_+(\mu)^L+\lambda_-(\mu)^L\right] \\
    &+  \frac{1}{2\pi}\int_0^{\pi} d\mu\left[\left(-\lambda_+(\mu)\right)^L+\left(-\lambda_-(\mu)\right)^L\right].
  \end{split}
    \label{ZIsing_cancellation}
\end{equation}
Equation (\ref{ZIsing_cancellation}) shows that for $L$ odd the partition function vanishes exactly. This can be readily understood by the impossibility of imposing the constraint $\sum_i S_i=0$ with an odd number of spin variables $S_i$, which take values $\pm 1$.
In the following we shall assume that $L$ is even, such that the two terms in Eq.~(\ref{ZIsing_cancellation}) are identical and we have
\begin{equation}
  Z_{\rm can} = \frac{1}{\pi}\int_0^{\pi} d\mu\left[\lambda_+(\mu)^L+\lambda_-(\mu)^L\right].
  \label{ZIsing_simplified}
\end{equation}
For large $L$ the integral of Eq.~(\ref{ZIsing_simplified}) is dominated by the saddle points of $\lambda_\pm(\mu)$ which are solutions of
\begin{equation}
\frac{d\lambda_\pm(\mu)}{d\mu} = e^{-K}\left(-\sin(\mu)\mp\frac{\sin(\mu)\cos(\mu)}{\sqrt{e^{4K}-\sin(\mu)^2}}\right)=0.
  \label{saddle}
\end{equation}
For both eigenvalues, Eq.~(\ref{saddle}) has solutions for $\mu=0$ and $\mu=\pi$, which lie at the border of the integration domain in Eq.~(\ref{ZIsing_simplified}). We observe that for $K>0$ (antiferromagnetic model), the eigenvalues are real. For $K<0$ (ferromagnetic model), $\lambda_\pm(\mu)$ given in Eq.~(\ref{Ising_lambda}) are real for $|\mu|<\varepsilon_0\equiv\arcsin(e^{2K})$ and $|\mu-\pi|<\varepsilon_0$, i.e., in an interval around the saddle points. For this reason, without losing generality it is convenient to shift the domain of integration in Eq.~(\ref{ZIsing_simplified})
\begin{equation}
  Z_{\rm can} = \frac{1}{\pi}\int_{-\varepsilon_0}^{\pi-\epsilon_0} d\mu\left[\lambda_+(\mu)^L+\lambda_-(\mu)^L\right],
  \label{ZIsing_shifted}
\end{equation}
where for $K>0$ one can take, e.g., $\varepsilon_0=\pi/2$, such that the single saddle point $\mu=0$ in the integration domain is an interior point. In order to determine the finite-size correction to the free energy, we need to compute the corrections around the saddle point. To this end, it is important to observe that for $K>0$ (antiferromagnetic model) both eigenvalues $\lambda_\pm(\mu)$ have a maximum around $\mu=0$; however, since $\lambda_-(\mu) < 0$, with $L$ even the term $\lambda_-(\mu)^L$ in Eq.~(\ref{ZIsing_shifted}) has a minimum at $\mu=0$, whereas $\lambda_+(\mu) > 0$ and $\lambda_+(\mu)^L$ has a maximum. For $K<0$ (ferromagnetic model) $\lambda_\pm(\mu)$ are real and positive in an interval around $\mu=0$; however, while $\lambda_+(\mu)$ has a maximum at $\mu=0$, the other eigenvalue $\lambda_-(\mu)$ has instead a minimum around $\mu=0$. Therefore, for both cases $K>0$ and $K<0$ it is not possible to separate Eq.~(\ref{ZIsing_shifted}) into a sum of two integrals to be evaluated for $L\rightarrow\infty$, but it is necessary to consider the behavior around $\mu=0$ of the sum of the two eigenvalues. To do so, we write the integrand of Eq.~(\ref{ZIsing_shifted}) as
\begin{equation}
  \begin{split}
&\lambda_+(\mu)^L+\lambda_-(\mu)^L = \exp\{Lg(\mu,L)\}, \\
&g(\mu,L)\equiv \ln\left[\left(\lambda_+(\mu)^L+\lambda_-(\mu)^L\right)^{1/L}\right].
  \end{split}
    \label{logeigen}
\end{equation}
A second-order Taylor expansion of $g(\mu,L)$ around $\mu=0$ gives
\begin{equation}
  \begin{split}
    g(\mu,L) &= \ln\left[\left(\lambda_+(0)^L+\lambda_-(0)^L\right)^{1/L}\right]\\
    &-\frac{e^{-2K}\left(1-\tanh(K)^L\right)}{2\left(1+\tanh(K)^L\right)}\mu^2+o(\mu^2).
  \end{split}
    \label{Taylorg}
\end{equation}
Equation (\ref{Taylorg}) shows that, indeed, $g(\mu,L)$ exhibits a maximum around $\mu=0$. Moreover, the coefficient in front of $\mu^2$ remains finite in the limit $L\rightarrow\infty$. Inserting the expansion of Eq.~(\ref{Taylorg}) in Eq.~(\ref{logeigen}), and using the resulting expression for $\lambda_+(\mu)^L+\lambda_-(\mu)^L$ in Eq.~(\ref{ZIsing_shifted}) we obtain, after a Gaussian integration,
\begin{equation}
  \begin{split}
    Z_{\rm can} \underset{L\rightarrow\infty}{\simeq} \frac{1}{\pi}&\left[\lambda_+(0)^L+\lambda_-(0)^L\right]\cdot\\
    &\left[\frac{2\pi\left(1+\tanh(K)^L\right)e^{2K}}{\left(1-\tanh(K)^L\right)L}\right]^{1/2}.
  \end{split}
  \label{ZIsing_largeL}
\end{equation}
The free energy per volume $L$, and in units of $k_BT$, $F_{\rm can}=-\ln Z_{\rm can}/L$ is
\begin{equation}
  \begin{split}
    F_{\rm can} \underset{L\rightarrow\infty}{\simeq} &\frac{1}{2L}\ln\left(\frac{\pi L}{2}\right) - \ln\lambda_+(0)  -\frac{K}{L}\\
    &-\frac{1}{L} \ln\left[1+\left(\frac{\lambda_-(0)}{\lambda_+(0)}\right)^L\right]\\
    &-\frac{1}{2L}\ln\left(\frac{1+\tanh(K)^L}{1-\tanh(K)^L}\right).
  \end{split}
  \label{FIsing_can}
\end{equation}
Using Eq.~(\ref{Ising_lambda}) and the known relation between the correlation length $\xi$ and the transfer-matrix eigenvalues
\begin{equation}
\xi = -\frac{1}{\ln \left|\lambda_-(0)/\lambda_+(0)\right|}=-\frac{1}{\ln\tanh |K|},
  \label{xi_Ising}
\end{equation}
Eq.~(\ref{FIsing_can}) can be written as
\begin{equation}
  \begin{split}
    F_{\rm can} &\underset{L\rightarrow\infty}{\simeq} \frac{1}{2L}\ln\left(\frac{\pi L}{2}\right) - \ln\left[2\cosh(K)\right]\\
    &\qquad -\frac{1}{L} \ln\left(1+e^{-L/\xi}\right) -\frac{1}{2L}\ln\left(\frac{1+e^{-L/\xi}}{1-e^{-L/\xi}}\right) -\frac{K}{L}\\
    &\simeq \frac{1}{2L}\ln\left(\frac{\pi L}{2}\right) - \ln\left[2\cosh(K)\right] -\frac{2}{L}e^{-L/\xi} -\frac{K}{L},
  \end{split}
  \label{FIsing_can_xi}
\end{equation}
where in the last equality we have expanded for $\xi/L\ll 1$, using the fact that $\xi$ is always finite.
Equation (\ref{FIsing_can_xi}) agrees with the general result of Eq.~(\ref{F_can_finitesize}), where subleading exponential finite-size corrections have been neglected. To confirm this, we observe that, under the mapping to the lattice gas model, the equivalent charge susceptibility $\chi_c$ (i.e., fluctuation of the particle number per volume and multiplied by $\beta$) is given by $\chi_c = \chi / 4$, where $\chi$ is the usual spin susceptibility which, for the one-dimensional Ising model, is $\chi=\beta \exp\{-2K\}$; on substituting $\chi_c\rightarrow \chi/4 = \beta \exp\{-2K\}/4$ in Eq.~(\ref{F_can_finitesize}) we recover Eq.~(\ref{FIsing_can_xi}). Alternatively, one can repeat the calculations of Sec.~\ref{sec:fscanonical}, fixing in Eq.~(\ref{Fcan}) the magnetization per volume $m$ instead of the filling fraction $n$. Then in the result of Eq.~(\ref{F_can_finitesize}), $\chi_c$ is replaced by the fluctuations of the magnetization, i.e., the usual spin susceptibility $\chi$. Moreover, different than for $n$, the allowed values for $m$ are separated by an interval of $2/V$. This results in a factor $1/2$ in front of the right-hand side of Eq.~(\ref{F_gc_continous}), which in turns gives rise to an additional contribution $-(\ln 2)/V$ to the right-hand side of Eq.~(\ref{F_can_finitesize}). Taking into account this additional term, and substituting $\chi_c\rightarrow\chi$ in Eq.~(\ref{F_can_finitesize}) we recover Eq.~(\ref{FIsing_can_xi}).

A comparison of Eq.~(\ref{FIsing_can_xi}) with the corresponding result for the grand-canonical ensemble
\begin{equation}
\begin{split}
  F_{\rm gc} &= - \ln\left[2\cosh(K)\right] -\frac{1}{L} \ln\left(1+e^{-L/\xi}\right)\\
  &\underset{L\rightarrow\infty}{\simeq} - \ln\left[2\cosh(K)\right] -\frac{1}{L}e^{-L/\xi}
\end{split}
  \label{FIsing_gcan}
\end{equation}
shows that, besides an irrelevant $L-$dependent constant, the free energy density in the canonical ensemble is affected by a finite $1/L$ correction to its thermodynamic limit $-\ln\left[2\cosh(K)\right]$, which is absent in the grand-canonical ensemble. We also notice that the constraint $M=0$ alters the coefficient in front of the subleading exponential correction $\exp\{-L/\xi\}$.
From Eq.~(\ref{FIsing_can_xi}) we can compute the energy density as
\begin{equation}
  E_{\rm can} = \frac{\partial F_{\rm can}}{\partial \beta} = -J\tanh(K) -\frac{J}{L} + O\left(e^{-L/\xi},\frac{1}{L^2}\right),
\label{EIsing_can}
\end{equation}
which exhibits a leading finite-size correction $\propto 1/L$. Due to the fact that $\chi/\beta$ is exactly exponential in the one-dimensional Ising model, such a finite-size term is temperature independent [see Eq.~(\ref{F_can_finitesize})].

As emphasized in the derivation of the results, Eq.~(\ref{FIsing_can_xi}) and Eq.~(\ref{EIsing_can}) are also valid for $J<0$, $K<0$, i.e., for a ferromagnetic model. We remark that it is not possible to take the limit $T\rightarrow 0$ in Eq.~(\ref{FIsing_can_xi}) and Eq.~(\ref{EIsing_can}) because the calculation assumes a finite correlation length $\xi$. Indeed, for $T\rightarrow 0$ the coefficient of $\mu^2$ in Eq.~(\ref{Taylorg}) either vanishes (for the antiferromagnetic model) or diverges in $L$ (for the ferromagnetic model), rendering the saddle-point expansion singular. In the ground state of the antiferromagnetic model $E_{\rm can} = -J$, with no size dependence.

\bibliographystyle{apsrev4-1_custom}
\bibliography{francesco,fassaad}

\begin{thebibliography}{27}%
\makeatletter
\providecommand \@ifxundefined [1]{%
 \@ifx{#1\undefined}
}%
\providecommand \@ifnum [1]{%
 \ifnum #1\expandafter \@firstoftwo
 \else \expandafter \@secondoftwo
 \fi
}%
\providecommand \@ifx [1]{%
 \ifx #1\expandafter \@firstoftwo
 \else \expandafter \@secondoftwo
 \fi
}%
\providecommand \natexlab [1]{#1}%
\providecommand \enquote  [1]{``#1''}%
\providecommand \bibnamefont  [1]{#1}%
\providecommand \bibfnamefont [1]{#1}%
\providecommand \citenamefont [1]{#1}%
\providecommand \href@noop [0]{\@secondoftwo}%
\providecommand \href [0]{\begingroup \@sanitize@url \@href}%
\providecommand \@href[1]{\@@startlink{#1}\@@href}%
\providecommand \@@href[1]{\endgroup#1\@@endlink}%
\providecommand \@sanitize@url [0]{\catcode `\\12\catcode `\$12\catcode
  `\&12\catcode `\#12\catcode `\^12\catcode `\_12\catcode `\%12\relax}%
\providecommand \@@startlink[1]{}%
\providecommand \@@endlink[0]{}%
\providecommand \url  [0]{\begingroup\@sanitize@url \@url }%
\providecommand \@url [1]{\endgroup\@href {#1}{\urlprefix }}%
\providecommand \urlprefix  [0]{URL }%
\providecommand \Eprint [0]{\href }%
\providecommand \doibase [0]{http://dx.doi.org/}%
\providecommand \selectlanguage [0]{\@gobble}%
\providecommand \bibinfo  [0]{\@secondoftwo}%
\providecommand \bibfield  [0]{\@secondoftwo}%
\providecommand \translation [1]{[#1]}%
\providecommand \BibitemOpen [0]{}%
\providecommand \bibitemStop [0]{}%
\providecommand \bibitemNoStop [0]{.\EOS\space}%
\providecommand \EOS [0]{\spacefactor3000\relax}%
\providecommand \BibitemShut  [1]{\csname bibitem#1\endcsname}%
\let\auto@bib@innerbib\@empty
\bibitem [{\citenamefont {{Barr{\'e}}}\ \emph {et~al.}(2001)\citenamefont
  {{Barr{\'e}}}, \citenamefont {{Mukamel}},\ and\ \citenamefont
  {{Ruffo}}}]{BMR-01}%
  \BibitemOpen
  \bibfield  {author} {\bibinfo {author} {\bibfnamefont {J.}~\bibnamefont
  {{Barr{\'e}}}}, \bibinfo {author} {\bibfnamefont {D.}~\bibnamefont
  {{Mukamel}}}, \ and\ \bibinfo {author} {\bibfnamefont {S.}~\bibnamefont
  {{Ruffo}}},\ }\bibfield  {title} {\bibinfo {title} {\emph {{Inequivalence of
  Ensembles in a System with Long-Range Interactions}}},\ }\href {\doibase
  10.1103/PhysRevLett.87.030601} {\bibfield  {journal} {\bibinfo  {journal}
  {\PRL}\ }\textbf {\bibinfo {volume} {87}},\ \bibinfo {eid} {030601} (\bibinfo
  {year} {2001})}\BibitemShut {NoStop}%
\bibitem [{\citenamefont {{Leyvraz}}\ and\ \citenamefont
  {{Ruffo}}(2002)}]{LR-02}%
  \BibitemOpen
  \bibfield  {author} {\bibinfo {author} {\bibfnamefont {F.}~\bibnamefont
  {{Leyvraz}}}\ and\ \bibinfo {author} {\bibfnamefont {S.}~\bibnamefont
  {{Ruffo}}},\ }\bibfield  {title} {\bibinfo {title} {\emph {{Ensemble
  inequivalence in systems with long-range interactions}}},\ }\href {\doibase
  10.1088/0305-4470/35/2/308} {\bibfield  {journal} {\bibinfo  {journal}
  {\JPAOLD}\ }\textbf {\bibinfo {volume} {35}},\ \bibinfo {pages} {285}
  (\bibinfo {year} {2002})}\BibitemShut {NoStop}%
\bibitem [{\citenamefont {Assaad}\ and\ \citenamefont
  {Evertz}(2008)}]{Assaad08_rev}%
  \BibitemOpen
  \bibfield  {author} {\bibinfo {author} {\bibfnamefont {F.}~\bibnamefont
  {Assaad}}\ and\ \bibinfo {author} {\bibfnamefont {H.}~\bibnamefont
  {Evertz}},\ }\bibfield  {title} {\bibinfo {title} {\emph {World-line and
  determinantal quantum monte carlo methods for spins, phonons and
  electrons}},\ }in\ \href {\doibase 10.1007/978-3-540-74686-7_10} {\emph
  {\bibinfo {booktitle} {Computational Many-Particle Physics}}},\ \bibinfo
  {series} {Lecture Notes in Physics}, Vol.\ \bibinfo {volume} {739},\ \bibinfo
  {editor} {edited by\ \bibinfo {editor} {\bibfnamefont {H.}~\bibnamefont
  {Fehske}}, \bibinfo {editor} {\bibfnamefont {R.}~\bibnamefont {Schneider}}, \
  and\ \bibinfo {editor} {\bibfnamefont {A.}~\bibnamefont {Wei{\ss}e}}}\
  (\bibinfo  {publisher} {Springer},\ \bibinfo {address} {Berlin Heidelberg},\
  \bibinfo {year} {2008})\ pp.\ \bibinfo {pages} {277--356}\BibitemShut
  {NoStop}%
\bibitem [{\citenamefont {{Koonin}}\ \emph {et~al.}(1997)\citenamefont
  {{Koonin}}, \citenamefont {{Dean}},\ and\ \citenamefont
  {{Langanke}}}]{KDK-97}%
  \BibitemOpen
  \bibfield  {author} {\bibinfo {author} {\bibfnamefont {S.~E.}\ \bibnamefont
  {{Koonin}}}, \bibinfo {author} {\bibfnamefont {D.~J.}\ \bibnamefont
  {{Dean}}}, \ and\ \bibinfo {author} {\bibfnamefont {K.}~\bibnamefont
  {{Langanke}}},\ }\bibfield  {title} {\bibinfo {title} {\emph {{Shell model
  Monte Carlo methods}}},\ }\href {\doibase 10.1016/S0370-1573(96)00017-8}
  {\bibfield  {journal} {\bibinfo  {journal} {\physrep}\ }\textbf {\bibinfo
  {volume} {278}},\ \bibinfo {pages} {1} (\bibinfo {year} {1997})}\BibitemShut
  {NoStop}%
\bibitem [{\citenamefont {{Bercx}}\ \emph {et~al.}(2017)\citenamefont
  {{Bercx}}, \citenamefont {{Goth}}, \citenamefont {{Hofmann}},\ and\
  \citenamefont {{Assaad}}}]{ALF}%
  \BibitemOpen
  \bibfield  {author} {\bibinfo {author} {\bibfnamefont {M.}~\bibnamefont
  {{Bercx}}}, \bibinfo {author} {\bibfnamefont {F.}~\bibnamefont {{Goth}}},
  \bibinfo {author} {\bibfnamefont {J.~S.}\ \bibnamefont {{Hofmann}}}, \ and\
  \bibinfo {author} {\bibfnamefont {F.~F.}\ \bibnamefont {{Assaad}}},\
  }\bibfield  {title} {\bibinfo {title} {\emph {{The ALF (Algorithms for
  Lattice Fermions) project release 1.0. Documentation for the auxiliary field
  quantum Monte Carlo code}}},\ }\href {\doibase 10.21468/SciPostPhys.3.2.013}
  {\bibfield  {journal} {\bibinfo  {journal} {SciPost Phys.}\ }\textbf
  {\bibinfo {volume} {3}},\ \bibinfo {pages} {013} (\bibinfo {year}
  {2017})}\BibitemShut {NoStop}%
\bibitem [{\citenamefont {{Bl{\"o}te}}\ \emph {et~al.}(2000)\citenamefont
  {{Bl{\"o}te}}, \citenamefont {{Heringa}},\ and\ \citenamefont
  {{Tsypin}}}]{BHT-00}%
  \BibitemOpen
  \bibfield  {author} {\bibinfo {author} {\bibfnamefont {H.~W.~J.}\
  \bibnamefont {{Bl{\"o}te}}}, \bibinfo {author} {\bibfnamefont {J.~R.}\
  \bibnamefont {{Heringa}}}, \ and\ \bibinfo {author} {\bibfnamefont {M.~M.}\
  \bibnamefont {{Tsypin}}},\ }\bibfield  {title} {\bibinfo {title} {\emph
  {{Three-dimensional Ising model in the fixed-magnetization ensemble: A Monte
  Carlo study}}},\ }\href {\doibase 10.1103/PhysRevE.62.77} {\bibfield
  {journal} {\bibinfo  {journal} {\pre}\ }\textbf {\bibinfo {volume} {62}},\
  \bibinfo {pages} {77} (\bibinfo {year} {2000})}\BibitemShut {NoStop}%
\bibitem [{\citenamefont {{Heringa}}\ and\ \citenamefont
  {{Bl{\"o}te}}(1998{\natexlab{a}})}]{HB-98}%
  \BibitemOpen
  \bibfield  {author} {\bibinfo {author} {\bibfnamefont {J.~R.}\ \bibnamefont
  {{Heringa}}}\ and\ \bibinfo {author} {\bibfnamefont {H.~W.~J.}\ \bibnamefont
  {{Bl{\"o}te}}},\ }\bibfield  {title} {\bibinfo {title} {\emph {{Geometric
  cluster Monte Carlo simulation}}},\ }\href {\doibase
  10.1103/PhysRevE.57.4976} {\bibfield  {journal} {\bibinfo  {journal} {\pre}\
  }\textbf {\bibinfo {volume} {57}},\ \bibinfo {pages} {4976} (\bibinfo {year}
  {1998}{\natexlab{a}})}\BibitemShut {NoStop}%
\bibitem [{\citenamefont {{Heringa}}\ and\ \citenamefont
  {{Bl{\"o}te}}(1998{\natexlab{b}})}]{HB-98B}%
  \BibitemOpen
  \bibfield  {author} {\bibinfo {author} {\bibfnamefont {J.~R.}\ \bibnamefont
  {{Heringa}}}\ and\ \bibinfo {author} {\bibfnamefont {H.~W.~J.}\ \bibnamefont
  {{Bl{\"o}te}}},\ }\bibfield  {title} {\bibinfo {title} {\emph {{Geometric
  symmetries and cluster simulations}}},\ }\href {\doibase
  10.1016/S0378-4371(98)00003-X} {\bibfield  {journal} {\bibinfo  {journal}
  {\PAA}\ }\textbf {\bibinfo {volume} {254}},\ \bibinfo {pages} {156} (\bibinfo
  {year} {1998}{\natexlab{b}})}\BibitemShut {NoStop}%
\bibitem [{\citenamefont {{L{\"u}scher}}(1986)}]{Luescher-86}%
  \BibitemOpen
  \bibfield  {author} {\bibinfo {author} {\bibfnamefont {M.}~\bibnamefont
  {{L{\"u}scher}}},\ }\bibfield  {title} {\bibinfo {title} {\emph {{Volume
  dependence of the energy spectrum in massive quantum field theories: I.
  Stable particle states}}},\ }\href {\doibase 10.1007/BF01211589} {\bibfield
  {journal} {\bibinfo  {journal} {Commun.~Math.~Phys.~}\ }\textbf {\bibinfo
  {volume} {104}},\ \bibinfo {pages} {177} (\bibinfo {year}
  {1986})}\BibitemShut {NoStop}%
\bibitem [{\citenamefont {{Neuberger}}(1989)}]{Neuberger-89}%
  \BibitemOpen
  \bibfield  {author} {\bibinfo {author} {\bibfnamefont {H.}~\bibnamefont
  {{Neuberger}}},\ }\bibfield  {title} {\bibinfo {title} {\emph {{Finite size
  effects in massive field theory}}},\ }\href {\doibase
  10.1016/0370-2693(89)90638-2} {\bibfield  {journal} {\bibinfo  {journal}
  {\PLB}\ }\textbf {\bibinfo {volume} {233}},\ \bibinfo {pages} {183} (\bibinfo
  {year} {1989})}\BibitemShut {NoStop}%
\bibitem [{\citenamefont {{M{\"u}nster}}(1985)}]{Muenster-85}%
  \BibitemOpen
  \bibfield  {author} {\bibinfo {author} {\bibfnamefont {G.}~\bibnamefont
  {{M{\"u}nster}}},\ }\bibfield  {title} {\bibinfo {title} {\emph {{The size of
  finite size effects in lattice gauge theories}}},\ }\href {\doibase
  10.1016/0550-3213(85)90027-6} {\bibfield  {journal} {\bibinfo  {journal}
  {\NPB}\ }\textbf {\bibinfo {volume} {249}},\ \bibinfo {pages} {659} (\bibinfo
  {year} {1985})}\BibitemShut {NoStop}%
\bibitem [{\citenamefont {{Montvay}}\ and\ \citenamefont
  {{Weisz}}(1987)}]{MW-87}%
  \BibitemOpen
  \bibfield  {author} {\bibinfo {author} {\bibfnamefont {I.}~\bibnamefont
  {{Montvay}}}\ and\ \bibinfo {author} {\bibfnamefont {P.}~\bibnamefont
  {{Weisz}}},\ }\bibfield  {title} {\bibinfo {title} {\emph {{Numerical study
  of finite volume effects in the 4-dimensional Ising model}}},\ }\href
  {\doibase 10.1016/0550-3213(87)90191-X} {\bibfield  {journal} {\bibinfo
  {journal} {\NPB}\ }\textbf {\bibinfo {volume} {290}},\ \bibinfo {pages} {327}
  (\bibinfo {year} {1987})}\BibitemShut {NoStop}%
\bibitem [{\citenamefont {{Cucchieri}}\ \emph {et~al.}(1997)\citenamefont
  {{Cucchieri}}, \citenamefont {{Mendes}}, \citenamefont {{Pelissetto}},\ and\
  \citenamefont {{Sokal}}}]{CMPS-97}%
  \BibitemOpen
  \bibfield  {author} {\bibinfo {author} {\bibfnamefont {A.}~\bibnamefont
  {{Cucchieri}}}, \bibinfo {author} {\bibfnamefont {T.}~\bibnamefont
  {{Mendes}}}, \bibinfo {author} {\bibfnamefont {A.}~\bibnamefont
  {{Pelissetto}}}, \ and\ \bibinfo {author} {\bibfnamefont {A.~D.}\
  \bibnamefont {{Sokal}}},\ }\bibfield  {title} {\bibinfo {title} {\emph
  {{Continuum limits and exact finite-size-scaling functions for
  one-dimensional O(N)-invariant spin models}}},\ }\href {\doibase
  10.1007/BF02199114} {\bibfield  {journal} {\bibinfo  {journal} {\JSF}\
  }\textbf {\bibinfo {volume} {86}},\ \bibinfo {pages} {581} (\bibinfo {year}
  {1997})}\BibitemShut {NoStop}%
\bibitem [{\citenamefont {{Caracciolo}}\ and\ \citenamefont
  {{Pelissetto}}(1998)}]{CP-98}%
  \BibitemOpen
  \bibfield  {author} {\bibinfo {author} {\bibfnamefont {S.}~\bibnamefont
  {{Caracciolo}}}\ and\ \bibinfo {author} {\bibfnamefont {A.}~\bibnamefont
  {{Pelissetto}}},\ }\bibfield  {title} {\bibinfo {title} {\emph {{Corrections
  to finite-size scaling in the lattice N-vector model for N=$\infty$}}},\
  }\href {\doibase 10.1103/PhysRevD.58.105007} {\bibfield  {journal} {\bibinfo
  {journal} {\prd}\ }\textbf {\bibinfo {volume} {58}},\ \bibinfo {eid} {105007}
  (\bibinfo {year} {1998})}\BibitemShut {NoStop}%
\bibitem [{\citenamefont {{Parisen Toldin}}\ \emph {et~al.}(2015)\citenamefont
  {{Parisen Toldin}}, \citenamefont {{Hohenadler}}, \citenamefont {{Assaad}},\
  and\ \citenamefont {{Herbut}}}]{PTHAH-14}%
  \BibitemOpen
  \bibfield  {author} {\bibinfo {author} {\bibfnamefont {F.}~\bibnamefont
  {{Parisen Toldin}}}, \bibinfo {author} {\bibfnamefont {M.}~\bibnamefont
  {{Hohenadler}}}, \bibinfo {author} {\bibfnamefont {F.~F.}\ \bibnamefont
  {{Assaad}}}, \ and\ \bibinfo {author} {\bibfnamefont {I.~F.}\ \bibnamefont
  {{Herbut}}},\ }\bibfield  {title} {\bibinfo {title} {\emph {{Fermionic
  quantum criticality in honeycomb and {$\pi$} -flux Hubbard models:
  Finite-size scaling of renormalization-group-invariant observables from
  quantum Monte Carlo}}},\ }\href {\doibase 10.1103/PhysRevB.91.165108}
  {\bibfield  {journal} {\bibinfo  {journal} {\prb}\ }\textbf {\bibinfo
  {volume} {91}},\ \bibinfo {eid} {165108} (\bibinfo {year}
  {2015})}\BibitemShut {NoStop}%
\bibitem [{\citenamefont {{Iyer}}\ \emph {et~al.}(2015)\citenamefont {{Iyer}},
  \citenamefont {{Srednicki}},\ and\ \citenamefont {{Rigol}}}]{ISR-15}%
  \BibitemOpen
  \bibfield  {author} {\bibinfo {author} {\bibfnamefont {D.}~\bibnamefont
  {{Iyer}}}, \bibinfo {author} {\bibfnamefont {M.}~\bibnamefont {{Srednicki}}},
  \ and\ \bibinfo {author} {\bibfnamefont {M.}~\bibnamefont {{Rigol}}},\
  }\bibfield  {title} {\bibinfo {title} {\emph {{Optimization of finite-size
  errors in finite-temperature calculations of unordered phases}}},\ }\href
  {\doibase 10.1103/PhysRevE.91.062142} {\bibfield  {journal} {\bibinfo
  {journal} {\pre}\ }\textbf {\bibinfo {volume} {91}},\ \bibinfo {eid} {062142}
  (\bibinfo {year} {2015})}\BibitemShut {NoStop}%
\bibitem [{\citenamefont {{Fukuda}}\ and\ \citenamefont
  {{Kyriakopoulos}}(1975)}]{FK-75}%
  \BibitemOpen
  \bibfield  {author} {\bibinfo {author} {\bibfnamefont {R.}~\bibnamefont
  {{Fukuda}}}\ and\ \bibinfo {author} {\bibfnamefont {E.}~\bibnamefont
  {{Kyriakopoulos}}},\ }\bibfield  {title} {\bibinfo {title} {\emph
  {{Derivation of the effective potential}}},\ }\href {\doibase
  10.1016/0550-3213(75)90014-0} {\bibfield  {journal} {\bibinfo  {journal}
  {\NPB}\ }\textbf {\bibinfo {volume} {85}},\ \bibinfo {pages} {354} (\bibinfo
  {year} {1975})}\BibitemShut {NoStop}%
\bibitem [{\citenamefont {{O'Raifeartaigh}}\ \emph {et~al.}(1986)\citenamefont
  {{O'Raifeartaigh}}, \citenamefont {{Wipf}},\ and\ \citenamefont
  {{Yoneyama}}}]{ORWY-86}%
  \BibitemOpen
  \bibfield  {author} {\bibinfo {author} {\bibfnamefont {L.}~\bibnamefont
  {{O'Raifeartaigh}}}, \bibinfo {author} {\bibfnamefont {A.}~\bibnamefont
  {{Wipf}}}, \ and\ \bibinfo {author} {\bibfnamefont {H.}~\bibnamefont
  {{Yoneyama}}},\ }\bibfield  {title} {\bibinfo {title} {\emph {{The constraint
  effective potential}}},\ }\href {\doibase 10.1016/S0550-3213(86)80031-1}
  {\bibfield  {journal} {\bibinfo  {journal} {\NPB}\ }\textbf {\bibinfo
  {volume} {271}},\ \bibinfo {pages} {653} (\bibinfo {year}
  {1986})}\BibitemShut {NoStop}%
\bibitem [{\citenamefont {{Palma}}(1992)}]{Palma-92}%
  \BibitemOpen
  \bibfield  {author} {\bibinfo {author} {\bibfnamefont {G.}~\bibnamefont
  {{Palma}}},\ }\bibfield  {title} {\bibinfo {title} {\emph {{Renormalized loop
  expansion to compute finite size effects of the constraint effective
  potential}}},\ }\href {\doibase 10.1007/BF01559498} {\bibfield  {journal}
  {\bibinfo  {journal} {\ZPC}\ }\textbf {\bibinfo {volume} {54}},\ \bibinfo
  {pages} {679} (\bibinfo {year} {1992})}\BibitemShut {NoStop}%
\bibitem [{\citenamefont {{Essam}}\ and\ \citenamefont
  {{Garelick}}(1967)}]{EG-67}%
  \BibitemOpen
  \bibfield  {author} {\bibinfo {author} {\bibfnamefont {J.~W.}\ \bibnamefont
  {{Essam}}}\ and\ \bibinfo {author} {\bibfnamefont {H.}~\bibnamefont
  {{Garelick}}},\ }\bibfield  {title} {\bibinfo {title} {\emph {{Critical
  behaviour of a soluble model of dilute ferromagnetism}}},\ }\href {\doibase
  10.1088/0370-1328/92/1/320} {\bibfield  {journal} {\bibinfo  {journal}
  {Proceedings of the Physical Society}\ }\textbf {\bibinfo {volume} {92}},\
  \bibinfo {pages} {136} (\bibinfo {year} {1967})}\BibitemShut {NoStop}%
\bibitem [{\citenamefont {{Fisher}}(1968)}]{Fisher-68}%
  \BibitemOpen
  \bibfield  {author} {\bibinfo {author} {\bibfnamefont {M.~E.}\ \bibnamefont
  {{Fisher}}},\ }\bibfield  {title} {\bibinfo {title} {\emph {{Renormalization
  of Critical Exponents by Hidden Variables}}},\ }\href {\doibase
  10.1103/PhysRev.176.257} {\bibfield  {journal} {\bibinfo  {journal} {\PR}\
  }\textbf {\bibinfo {volume} {176}},\ \bibinfo {pages} {257} (\bibinfo {year}
  {1968})}\BibitemShut {NoStop}%
\bibitem [{\citenamefont {Fisher}\ and\ \citenamefont
  {de~Gennes}(1978)}]{FG-78}%
  \BibitemOpen
  \bibfield  {author} {\bibinfo {author} {\bibfnamefont {M.~E.}\ \bibnamefont
  {Fisher}}\ and\ \bibinfo {author} {\bibfnamefont {P.-G.}\ \bibnamefont
  {de~Gennes}},\ }\bibfield  {title} {\bibinfo {title} {\emph {{Ph\'enom\`enes
  aux parois dans un m\'elange binaire critique}}},\ }\href
  {http://gallica.bnf.fr/ark:/12148/bpt6k62353730/f61.image.r=fisher%20de%20gennes.langEN}
  {\bibfield  {journal} {\bibinfo  {journal} {C.~R.~Acad.~Sci.~Paris Ser.~B~}\
  }\textbf {\bibinfo {volume} {287}},\ \bibinfo {pages} {207} (\bibinfo {year}
  {1978})}\BibitemShut {NoStop}%
\bibitem [{\citenamefont {{Gross}}\ \emph {et~al.}(2016)\citenamefont
  {{Gross}}, \citenamefont {{Vasilyev}}, \citenamefont {{Gambassi}},\ and\
  \citenamefont {{Dietrich}}}]{GVGD-16}%
  \BibitemOpen
  \bibfield  {author} {\bibinfo {author} {\bibfnamefont {M.}~\bibnamefont
  {{Gross}}}, \bibinfo {author} {\bibfnamefont {O.}~\bibnamefont {{Vasilyev}}},
  \bibinfo {author} {\bibfnamefont {A.}~\bibnamefont {{Gambassi}}}, \ and\
  \bibinfo {author} {\bibfnamefont {S.}~\bibnamefont {{Dietrich}}},\ }\bibfield
   {title} {\bibinfo {title} {\emph {{Critical adsorption and critical Casimir
  forces in the canonical ensemble}}},\ }\href {\doibase
  10.1103/PhysRevE.94.022103} {\bibfield  {journal} {\bibinfo  {journal}
  {\pre}\ }\textbf {\bibinfo {volume} {94}},\ \bibinfo {eid} {022103} (\bibinfo
  {year} {2016})}\BibitemShut {NoStop}%
\bibitem [{\citenamefont {Ormand}\ \emph {et~al.}(1994)\citenamefont {Ormand},
  \citenamefont {Dean}, \citenamefont {Johnson}, \citenamefont {Lang},\ and\
  \citenamefont {Koonin}}]{Ormand94}%
  \BibitemOpen
  \bibfield  {author} {\bibinfo {author} {\bibfnamefont {W.~E.}\ \bibnamefont
  {Ormand}}, \bibinfo {author} {\bibfnamefont {D.~J.}\ \bibnamefont {Dean}},
  \bibinfo {author} {\bibfnamefont {C.~W.}\ \bibnamefont {Johnson}}, \bibinfo
  {author} {\bibfnamefont {G.~H.}\ \bibnamefont {Lang}}, \ and\ \bibinfo
  {author} {\bibfnamefont {S.~E.}\ \bibnamefont {Koonin}},\ }\bibfield  {title}
  {\bibinfo {title} {\emph {Demonstration of the auxiliary-field monte carlo
  approach for sd-shell nuclei}},\ }\href {\doibase 10.1103/PhysRevC.49.1422}
  {\bibfield  {journal} {\bibinfo  {journal} {Phys. Rev. C}\ }\textbf {\bibinfo
  {volume} {49}},\ \bibinfo {pages} {1422} (\bibinfo {year}
  {1994})}\BibitemShut {NoStop}%
\bibitem [{\citenamefont {Gilbreth}\ and\ \citenamefont
  {Alhassid}(2015)}]{Gilbreth15}%
  \BibitemOpen
  \bibfield  {author} {\bibinfo {author} {\bibfnamefont {C.}~\bibnamefont
  {Gilbreth}}\ and\ \bibinfo {author} {\bibfnamefont {Y.}~\bibnamefont
  {Alhassid}},\ }\bibfield  {title} {\bibinfo {title} {\emph {Stabilizing
  canonical-ensemble calculations in the auxiliary-field monte carlo method}},\
  }\href {\doibase http://dx.doi.org/10.1016/j.cpc.2014.09.002} {\bibfield
  {journal} {\bibinfo  {journal} {Computer Physics Communications}\ }\textbf
  {\bibinfo {volume} {188}},\ \bibinfo {pages} {1 } (\bibinfo {year}
  {2015})}\BibitemShut {NoStop}%
\bibitem [{\citenamefont {{Gross}}\ \emph {et~al.}(2017)\citenamefont
  {{Gross}}, \citenamefont {{Gambassi}},\ and\ \citenamefont
  {{Dietrich}}}]{GGD-17}%
  \BibitemOpen
  \bibfield  {author} {\bibinfo {author} {\bibfnamefont {M.}~\bibnamefont
  {{Gross}}}, \bibinfo {author} {\bibfnamefont {A.}~\bibnamefont {{Gambassi}}},
  \ and\ \bibinfo {author} {\bibfnamefont {S.}~\bibnamefont {{Dietrich}}},\
  }\bibfield  {title} {\bibinfo {title} {\emph {{Statistical field theory with
  constraints: application to critical Casimir forces in the canonical
  ensemble}}},\ }\href {\doibase 10.1103/PhysRevE.96.022135} {\bibfield
  {journal} {\bibinfo  {journal} {\PRE}\ }\textbf {\bibinfo {volume} {96}},\
  \bibinfo {pages} {022135} (\bibinfo {year} {2017})}\BibitemShut {NoStop}%
\bibitem [{\citenamefont {Krause}\ and\ \citenamefont
  {Th\"ornig}(2016)}]{Jureca16}%
  \BibitemOpen
  \bibfield  {author} {\bibinfo {author} {\bibfnamefont {D.}~\bibnamefont
  {Krause}}\ and\ \bibinfo {author} {\bibfnamefont {P.}~\bibnamefont
  {Th\"ornig}},\ }\bibfield  {title} {\bibinfo {title} {\emph {{JURECA:
  General-purpose supercomputer at J\"ulich Supercomputing Centre}}},\ }\href
  {http://dx.doi.org/10.17815/jlsrf-2-121} {\bibfield  {journal} {\bibinfo
  {journal} {Journal of large-scale research facilities}\ }\textbf {\bibinfo
  {volume} {2}},\ \bibinfo {pages} {A62} (\bibinfo {year} {2016})}\BibitemShut
  {NoStop}%
\end{thebibliography}%
\end{document}